%% file: paper.tex
\documentclass[fleqn]{2023SCGE}

\usepackage[utf8]{inputenc}
\usepackage{soul}
\input{figures}

\input{tables}

\begin{document}

\ArticleType{Article}

\Year{2026}
\Month{}
\Vol{}
\No{}
\DOI{}
\BeginPage{1} 
\EndPage{}
\ReceiveDate{Month Day, Year}
\AcceptDate{Month Day, Year}
\OnlineDate{Month Day, Year}

\title{Forecasting cosmological constraints from the nonlinear halo density field with Zel’dovich-inspired transform}
{Constraining cosmology from the halo field with ZI transform}

\author[1]{Yu Zhou}{}
\author[1]{Yun Wang}{yunw@jlu.edu.cn}
\author[1]{Haozhe Zhang}{}
\author[2]{Matteo Peron}{}
\author[3,4]{Shiyu Deng}{dengsy@cho.ac.cn}%
\author[5,6,7]{Yu Yu}{}
\author[8,9,10]{Xin Zhang}{zhangxin@neu.edu.cn}

\AuthorMark{Zhou Y}

\AuthorCitation{Zhou Y, Wang Y, Zhang H, et al}

\address[1]{Key Laboratory of Material Simulation Methods \& Software of Ministry of Education, College of Physics, Jilin University, Changchun {\rm 130012}, China}
\address[2]{Dipartimento di Scienze Matematiche, Fisiche e Informatiche, Universit\`{a} di Parma, and INFN Gruppo Collegato di Parma, \\ Parco Area delle Scienze 7/A, I-43124, Parma, Italy}
\address[3]{Changchun Observatory of National Astronomical Observators, Chinese Academy of Sciences, Changchun {\rm 130117}, China}
\address[4]{National Astronomical Observators, Chinese Academy of Sciences, Beijing {\rm100101}, China}
\address[5]{State Key Laboratory of Dark Matter Physics, Shanghai Jiao Tong University, Shanghai {\rm200240}, China}
\address[6]{Department of Astronomy, School of Physics and Astronomy, Shanghai Jiao Tong University, Shanghai {\rm200240}, China}
\address[7]{Key Laboratory for Particle Astrophysics and Cosmology (MOE)/Shanghai Key Laboratory for Particle Physics and Cosmology, \\ Shanghai Jiao Tong University,
Shanghai {\rm200240}, China}
\address[8]{Liaoning Key Laboratory of Cosmology and Astrophysics \& College of Sciences, Northeastern University, Shenyang {\rm 110819}, China}
\address[9]{Key Laboratory of Data Analytics and Optimization for Smart Industry (Ministry of Education), Northeastern University, Shenyang {\rm 110819}, China}
\address[10]{National Frontiers Science Center for Industrial Intelligence and Systems Optimization, Northeastern University, Shenyang {\rm 110819}, China}


\abstract{
The Zel’dovich-inspired (ZI) transform is a local nonlinear transform of the density field that can Gaussianize the late-time matter field and recover information from higher-order $N$-point correlation functions through its power spectrum, with the parameter $\eta$ controlling the weighting of higher-order information. Using Fisher-matrix analyses of halo catalogs from the Quijote and Quijote-PNG simulations, we study the redshift-space monopole power spectrum of the ZI-transformed halo density field. The joint statistic $P_{\rm ZI,0}$, constructed from three representative transforms with $\eta=1$, $10$, and $+\infty$, yields substantially tighter constraints than the conventional halo power spectrum $P_{\rm h,0}$ on standard cosmological parameters, the summed neutrino mass, and primordial non-Gaussianity amplitudes, while outperforming the marked power spectrum for many parameters. We further calibrate a scale-dependent noise correction for the ZI power spectrum, with an effective noise amplitude that depends on wavenumber, and find that the correction further tightens the constraints. In particular, the noise-corrected $P_{\rm ZI,0}$ improves the constraints on $f_{\rm NL}^{\rm equil}$, $f_{\rm NL}^{\rm ortho}$, $M_\nu$, and $n_{\rm s}$ by factors of $5.6$, $5.5$, $5.2$, and $5.8$, respectively, relative to $P_{\rm h,0}$. These results demonstrate the potential of ZI power spectra as a complement to standard two-point statistics for extracting nonlinear cosmological information. Our forecasts may be conservative because the halo catalogs have substantially lower number densities than those expected from forthcoming galaxy surveys.
}


\maketitle



\begin{multicols}{2}

\section{Introduction}\label{sec:intro}

The $\Lambda$CDM model provides the standard framework for describing the evolution of the Universe, which is characterized by a set of cosmological parameters, including the matter density $\Omega_{\rm m}$, baryon density $\Omega_{\rm b}$, the Hubble parameter $h$, the scalar spectral index $n_{\rm s}$, and the amplitude of density fluctuations $\sigma_8$. Extensions of the standard model also introduce additional parameters, such as the sum of neutrino masses $M_\nu \equiv \sum m_{\nu_i}$, which plays an important role in cosmic structure formation through the free-streaming effect of massive neutrinos. Beyond these cosmological parameters, primordial non-Gaussianity (PNG), commonly parameterized by $f_{\rm NL}$, provides a unique probe of the physics of the early Universe. While the simplest single-field slow-roll inflation models predict $\left|f_{\rm NL}\right|\ll1$ \cite{Wang_2000,Maldacena_2003}, many alternative inflationary models predict observable levels of PNG \cite{Lyth_2003,Alishahiha_2004,Bartolo_2004}, placing tight constraints on $f_{\rm NL}$ provides an powerful test of inflationary models.

Over the past two decades, high-precision observations of the cosmic microwave background (CMB) have yielded tight constraints on cosmological parameters \cite{Hinshaw_2013,Planck2020,Aiola_2020,Ade_2021,Balkenhol_2023,Jung_2025}. However, the CMB primarily probes the two-dimensional surface of last scattering, which limits the number of independent Fourier modes available for inference and ultimately renders its constraining power limited by cosmic variance \cite{Kalaja_2021}. Compared with the CMB, large-scale structure (LSS) surveys provide a three-dimensional view of the Universe, encoding information about the formation and subsequent evolution of cosmic structures across a broad range of scales and redshifts. The LSS contains a wealth of complementary information that can be traced by different observables and tracers. Jointly exploiting these multiple tracers can extract more of the available cosmological information while mitigating cosmic variance and breaking parameter degeneracies, thereby improving the precision and accuracy of cosmological parameter constraints \cite{Seljak_2009,barreira2023optimalrobustfrmnl,Karagiannis_2024}.

The fourth-generation galaxy surveys, such as DESI \cite{DESI_2016}, Euclid \cite{Euclid2025}, LSST \cite{LSST_2019}, and the forthcoming CSST \cite{CSST2026}, will map the LSS of the Universe with unprecedented precision over vast survey volumes. However, as cosmic structures evolve into the non-linear regime, the matter density field progressively deviates from Gaussianity, causing an increasing fraction of the cosmological information to be encoded in higher-order $N$-point correlation functions. To fully harness this non-linear information, a variety of analysis techniques have been developed. These include, for example, reconstructing real-space density fields from redshift-space galaxy observations \cite{Wang_2020_Iterative,chen2024,Shi_2025}, reconstructing the initial density field to mitigate the effects of late-time non-linear gravitational evolution \cite{Wang_2024_Extracting,Fl_ss_2024,Nguyen_2024,bayer_2026}, and machine-learning-based field-level inference \cite{He_2019,Villaescusa_Navarro_2021,lemos_2023}. Other approaches focus on constructing novel observables and statistical measures, such as scale-dependent void bias \cite{Chan_2014,Chan_2019,Schuster_2019}, marked power spectra \cite{Massara_2023,Cowell_2024,massara_2024,ebina2026}, density-split statistics \cite{Friedrich_2018,Paillas_2023,paillas_2023_BOSS}, cosmic web dependent statistics \cite{Bonnaire_2022,Bonnaire_2023,Sunseri_2025}, and wavelet-based statistics \cite{Cheng2020MNRAS,Wang_2022_Simultaneous,peron_2024,Wang_2025}. Although these methods have improved cosmological parameter constraints to varying degrees, they may involve substantial computational costs, require complex theoretical modeling, or face challenges in interpretability. Consequently, developing statistical methods that can efficiently extract cosmological information while maintaining computational tractability remains an important direction in modern cosmology.

Among various approaches, local non-linear transforms of the density field provide an efficient and computationally simple framework for extracting cosmological information from non-linear LSS. These transforms, including the logarithmic transform \cite{Neyrinck_2009,Carron_2013,Wang_2025}, the Box-Cox transform \cite{Joachimi_2011_Forecasts,Joachimi_2011_Cosmological,Schuhmann_2016}, and the clipping transform \cite{Simpson_2011,Simpson_2013}, apply local mappings to suppress high-density fluctuations and enhance underdense regions, thereby making the transformed field more linear and Gaussian. Previous studies have demonstrated that the power spectra of transformed fields can bring back information from higher-order $N$-point correlation functions, improving constraints on cosmological parameters \cite{Neyrinck_2009,wolk2014,Hall_2017,Wang_2025}. Compared with directly measuring higher-order $N$-point correlation functions such as the bispectrum and trispectrum, local transforms offer significant computational advantages, as they require only a point-wise transform followed by standard power spectrum analysis, avoiding the explicit evaluation of the large number of wavevector configurations \cite{Bertolini_2016,Watkinson_2017}. Thus, local non-linear transforms provide an efficient alternative for extracting non-Gaussian information from large-scale structure while maintaining practical feasibility for cosmological inference.

In our previous work, we introduced the ZI transform \cite{wang2026}, a single-parameter transform in which $\eta$ controls the weighting of higher-order information in the power spectrum of the transformed field. For $\eta \geq 3$, the one-point probability distribution of the transformed density field is approximately Gaussian, with the highest degree of Gaussianity attained at $\eta = 6$. We applied this transform to dark matter density fields from the Quijote simulations \cite{Villaescusa_Navarro_2020} and Quijote-PNG simulations \cite{Coulton_2023}, and employed Fisher information analysis to assess its constraining power on standard cosmological parameters, the summed neutrino mass, and amplitudes of different PNG shapes. We found that the power spectra of the transformed fields can significantly improve parameter constraints compared with the standard dark matter power spectrum. And, combining power spectra measured with multiple $\eta$ values allows complementary information from different orders of the underlying density statistics to be captured in the power spectrum, enhancing the overall constraining power. This multi-$\eta$ approach can achieve more competitive performance than the bispectrum and several other advanced summary statistics for most parameters, while retaining a simple and computationally efficient construction that is straightforward and fast to evaluate from the density field. The ZI transform is also particularly relevant for massive neutrinos, as neutrinos induce scale-dependent modifications to the halo--matter connection, which can help break the degeneracy between the neutrino mass and other parameters such as $\sigma_8$ \cite{Chiang_2019}. The local nature of the ZI transform makes it well suited to capturing such scale-dependent signatures directly from the power spectrum, without requiring higher-order statistics. These results motivate extending the ZI transform to more realistic tracers of LSS. Dark matter halos provide a natural intermediate step toward observed galaxies, but their density fields introduce additional effects, including nonlinear halo bias \cite{Desjacques_2018}, stochastic noise \cite{Hamaus_2010}, and mass-dependent clustering \cite{Dalal_2008}. Quantifying the impact of these effects is therefore important for assessing the applicability of ZI-transformed halo statistics to future galaxy surveys.

In this work, we extend the ZI transform from continuous matter fields to halo density fields in redshift space and investigate its capability to extract cosmological information from realistic tracers. We account for the dependence of the transformed halo power spectra on tracer bias by varying the minimum halo mass threshold, $M_{\min}$, around the fiducial value of $3.2\times10^{13},h^{-1}M_\odot$. Following \cite{Coulton_2023b}, we use halo catalogs with $M_{\min}=3.1$ and $3.3\times10^{13},h^{-1}M_\odot$, processed in an identical manner, to estimate the response to the halo mass threshold and hence to tracer bias. We then evaluate the resulting cosmological constraints using Fisher analysis. Furthermore, we propose a treatment to reduce the impact of noise in the ZI-transformed halo fields arising from the discrete sampling of halos. Our findings demonstrate the potential of ZI-transformed halo statistics as an efficient probe for future large-scale structure surveys.

The paper is organized as follows. Section \ref{sec:zi_trans} introduces the ZI transform and the proposed  noise correction method. Section \ref{sec:simulation} describes the N-body simulations and halo catalogs used in this work. Section \ref{sec:fisher} presents the Fisher formalism adopted to quantify the cosmological information extracted from different statistical measurements. Section \ref{sec:results} presents the main results of the Fisher analysis and discusses their implications. Finally, Section \ref{sec:conclusions} summarizes our findings and provides concluding remarks.

\section{ZI transform for halo density fields}
\label{sec:zi_trans}

In this work, we apply the ZI transform to halo density fields. By introducing the high-order spectral-weighting parameter $\eta$, the ZI transform controls the strength of the nonlinear mapping, allowing us to investigate the impact of different transform levels on the cosmological information extracted from halo fields. We further develop a  noise correction method to mitigate the contribution of  noise in the transformed halo density fields.

\subsection{Definition of ZI transform}

For the halo density field, we first define the halo overdensity field as
$\delta_h(\bm{x})
=
\frac{n(\bm{x})}{\bar n}-1,$
where $n(\boldsymbol{x})$ denotes the halo number density at position $\boldsymbol{x}$, and $\bar n$ is the mean halo number density. The ZI transform is defined as
\begin{equation}
\delta_{\mathrm{ZI}-\eta}
\equiv
\eta
\left[
1-(1+\delta_h)^{-1/\eta}
\right],
\label{eq:zi_transform}
\end{equation}
where the parameter $\eta$ controls the strength of the transform: overdense regions are compressed toward the upper bound $\delta_{\mathrm{ZI}-\eta}\to\eta$, while underdense regions are stretched to more negative values. 
This behavior is clearly illustrated in Figure~\ref{fig:ZI}, where the compression of overdense regions and the stretching of underdense regions become increasingly pronounced as $\eta$ decreases. In the limit $\eta\rightarrow+\infty$, the ZI transform approaches the logarithmic transform,
\begin{equation}
\delta_{\mathrm{log}}
\equiv
\ln(1+\delta_h).
\label{eq:log_transform}
\end{equation}

\figZI

For a discrete halo density field, sparse halo sampling naturally produces empty grid cells with $\delta_h=-1$. Consequently, Eqs.~\eqref{eq:zi_transform} and \eqref{eq:log_transform} cannot be applied directly to such cells, as both transforms become singular at $1+\delta_h=0$. To avoid the numerical divergence at $\delta_h=-1$, we introduce an artificial floor for extremely low-density regions. Specifically, for grid cells satisfying
$1+\delta_h < 10^{-4}$,
we set
$1+\delta_h = 0.5\times10^{-4}$.
After this treatment, the minimum halo overdensity is
$\delta_h^{\rm min}=-0.99995$,
thereby ensuring that the ZI transform remains finite throughout the calculation.

Then, the power spectrum of the ZI-transformed field is defined through
\begin{equation}
\left\langle
\delta_{\mathrm{ZI}-\eta}(\bm{k})
\delta_{\mathrm{ZI}-\eta}^{*}(\bm{k}')
\right\rangle
=
(2\pi)^3
\delta_{\mathrm{D}}(\bm{k}-\bm{k}')
P_{\mathrm{ZI}-\eta}(k),
\label{eq:zi_power_spectrum}
\end{equation}
where $P_{\mathrm{ZI}-\eta}(k)$ denotes the power spectrum of the ZI-transformed halo density field.

\subsection{Noise correction}

For the power spectra of any local non-linear transformed halo fields, there is no universal analytical prescription for correcting the  noise contribution. McCullagh et al. \cite{McCullagh_2016} proposed an empirical model for the power spectrum of nonlinearly transformed fields, where the measured power spectrum is expressed as
$
P^{\rm raw}(k|\bar n)
=
P^{\rm corrected}(k)
+
A/\bar{n}
$,
where $A$ is a constant parameter describing the noise contribution. 

\fignoise

In this work, we generalize this approach by assuming that the noise coefficient has a scale dependence, i.e., $A\equiv A(k|\eta)$. The power spectrum of the ZI-transformed halo field can therefore be written as
$
P^\mathrm{raw}_{\mathrm{ZI}-\eta}(k|\bar n)
=
P^\mathrm{corrected}_{\mathrm{ZI}-\eta}(k)
+
A(k|\eta)/\bar{n}
$.

By measuring the power spectra from two halo catalogs with different mean number densities, $\bar n_1$ and $\bar n_2$, the scale-dependent noise contribution can be estimated as
\begin{equation}
A(k|\eta)=
\frac{
P^\mathrm{raw}_{\mathrm{ZI}-\eta}(k|\bar n_1)
-
P^\mathrm{raw}_{\mathrm{ZI}-\eta}(k|\bar n_2)
}
{
1/\bar n_1-1/\bar n_2
}.
\end{equation}

The original halo catalog has a mean number density $\bar n_0$. We generate additional halo catalogs through random downsampling, with mean number densities $\bar n_1=0.94\,\bar n_0, \bar n_2=0.88\,\bar n_0, \bar n_3=0.82\,\bar n_0$. The downsampling procedure changes only the sampling density while preserving the underlying clustering signal. Therefore, four halo catalogs with different number densities allow us to obtain six independent estimates of $A_i(k|\eta)$ from the catalog pairs $(\bar n_0,\bar n_1)$, $(\bar n_0,\bar n_2)$, $(\bar n_0,\bar n_3)$, $(\bar n_1,\bar n_2)$, $(\bar n_1,\bar n_3)$, and $(\bar n_2,\bar n_3)$. We then average these estimates to reduce the statistical uncertainty:
$
A(k|\eta)=\frac{1}{6}\sum_{i=0}^{5}A_i(k|\eta)
$. Finally, the noise-corrected power spectrum of the ZI-transformed halo field is obtained as
\begin{equation}
P^\mathrm{corrected}_{\mathrm{ZI}-\eta}(k)
=
P^\mathrm{raw}_{\mathrm{ZI}-\eta}(k|\bar n_0)
-
\frac{A(k|\eta)}{\bar n_0}.
\end{equation}

As shown in Figure~\ref{fig:noise}, we test the noise correction using the original halo catalog and three randomly downsampled catalogs with mean number densities $\bar n=0.94\bar n_0$, $0.88\bar n_0$, and $0.82\bar n_0$, respectively. The upper panel shows the raw $P_{\rm ZI-1}$ power spectra, together with the corresponding scale-dependent noise contributions $-A(k|\eta)/\bar n$, where $A(k|\eta)/\bar n$ is negative and thus $-A(k|\eta)/\bar n$ is positive, and the noise-corrected power spectra. The lower panel shows the ratios of the corrected power spectra from the four catalogs to their mean power spectrum. Despite the different sampling number densities, the corrected power spectra are highly consistent with one another, demonstrating that the noise correction effectively removes the dependence on the sampling number density. Although only the case of $\eta=1$ is shown here, we find the same behavior for other values of $\eta$. As shown in the upper panel, the scale dependence of the noise contribution closely resembles that of the noise-corrected ZI power spectrum, $P^\mathrm{corrected}_{\mathrm{ZI}-\eta}(k)$. This similarity may provide useful guidance for future attempts to develop a theoretical model for $A(k|\eta)$.

\subsection{Choice of the $\eta$ values}

To determine suitable values of the $\eta$ parameter, we compare the shape of the transformed power spectra for different $\eta$ values with that of the initial linear matter power spectrum. Since the ZI power spectra corresponding to different $\eta$ values have different overall amplitudes, we first normalize the ZI power spectra on large scales before comparing their shapes. Specifically, we define
$
b_0
=
P_{\mathrm{ZI}-\eta}(k_0)
/
P_{\mathrm{initial}}(k_0)
$,
where $k_0$ denotes the minimum wavenumber considered. The quantity $b_0$ characterizes the overall large-scale amplitude ratio between the ZI-transformed power spectrum and the initial linear matter power spectrum. Dividing the ZI power spectrum by $b_0$ therefore removes the overall amplitude difference associated with different $\eta$ values, allowing us to focus primarily on differences in the spectral shape.

We then define a quantity to quantify the deviation of the transformed power spectrum from the initial linear power spectrum:
\begin{equation}
\Delta_\eta
=
\frac{1}{N_k}
\sum_{k<0.5}
\left|
\frac{
P_{\mathrm{ZI}-\eta}(k)/b_0
-
P_{\mathrm{initial}}(k)
}{
P_{\mathrm{initial}}(k)
}
\right|,
\end{equation}
where $N_k$ is the number of $k$ bins satisfying $k<0.5\,h\,\mathrm{Mpc}^{-1}$. The quantity $\Delta_\eta$ represents the mean relative deviation of the normalized ZI power spectrum from the initial linear matter power spectrum over the considered range of wavenumbers. A smaller value of $\Delta_\eta$ therefore indicates greater similarity between the shapes of the transformed and initial linear power spectra.

\figeta

Figure~\ref{fig:eta} shows the dependence of $\Delta_\eta$ on $\eta$. It can be seen that the noise-corrected power spectra generally exhibit smaller values of $\Delta_\eta$ across the considered range of $\eta$. This indicates that noise correction makes the shape of the transformed power spectrum closer to that of the initial linear matter power spectrum. The figure also shows that $\Delta_\eta$ generally increases with increasing $\eta$, indicating that the shape of the transformed power spectrum gradually deviates from that of the initial linear power spectrum. On the other hand, smaller values of $\eta$ lead to a stronger response of the ZI transform in underdense regions, which may enhance numerical uncertainties associated with insufficient sampling in low-density regions. Therefore, the choice of $\eta$ should take into account the preservation of the power-spectrum shape, numerical stability, and the information complementarity among different transforms.

\tabsimulations

Considering these factors together, we adopt $\eta=1, 10$ and $+\infty(\log)$, as the three representative choices used throughout our analysis. These choices span three representative cases with distinct values of $\Delta_\eta$, while avoiding $\eta$ values that are too close to each other, thereby providing greater information complementarity among the resulting transformed power spectra.

\section{Simulation sets}\label{sec:simulation}

In this work, we use the publicly available Quijote \cite{Villaescusa_Navarro_2020} and Quijote-PNG \cite{Coulton_2023} N-body simulation suites to construct halo number density fields. All halo catalogs are taken from the $z=0$ snaps.

\subsection{N-body Simulations}

The Quijote simulation suite consists of more than 44,000 $N$-body simulations, performed in periodic cubic volumes with side length $L=1\,h^{-1}{\rm Gpc}$ using the TreePM code Gadget-III, a later version of the publicly available Gadget-II code \cite{Springel_2005}. The initial conditions (ICs) are generated at redshift $z=127$ using second-order Lagrangian Perturbation Theory (2LPT) \cite{Jenkins_2010} for most cosmological models, while the Zel'dovich Approximation (ZA) \cite{White_2014} is adopted for massive-neutrino simulations. The Quijote-PNG suite extends Quijote to models with PNG.

\figcorrelation

In this work, the simulations are divided into two groups according to their roles in the analysis. First, we use 14,500 realizations from the fiducial Quijote simulations, generated with 2LPT initial conditions and adopting the Planck 2018 cosmology \cite{Planck2020}, to estimate the covariance matrix of the observables. The remaining simulation sets are used to compute numerical derivatives with respect to model parameters through finite differences. Specifically, these include: (i) the PNG simulations from Quijote-PNG, generated with 2LPTPNG \cite{Scoccimarro_2012} initial conditions, consisting of the local, equilateral, and orthogonal shapes; (ii) the massive-neutrino simulations ($M_\nu=0.1$, $0.2$, and $0.4\,\mathrm{eV}$) together with their corresponding fiducial\_ZA simulations ($M_\nu=0\,\mathrm{eV}$), all generated using ZA initial conditions to ensure consistency in the finite-difference calculation of neutrino-mass derivatives; (iii) the standard cosmological parameter variation simulations, in which small positive and negative perturbations are applied to the fiducial values of $h$, $n_{\rm s}$, $\Omega_{\rm m}$, $\Omega_{\rm b}$, and $\sigma_8$ using 2LPT initial conditions; and (iv) two additional halo catalogs with different minimum halo mass thresholds, constructed from a reserved subset of 500 fiducial realizations, which are used to evaluate the response of the observables to variations in the halo mass threshold. The detailed parameter settings for each simulation set are summarized in Table~\ref{tab:simulations}.

\subsection{Halo Catalogs}

Dark matter halos in the Quijote simulations are identified using the Friends-of-Friends (FoF) algorithm \cite{More_2011} with a linking length parameter of $b=0.2$, where particles separated by distances smaller than the linking length are grouped into the same halo.

To construct the halo samples, we select halos according to their FoF masses. For the majority of the simulation sets, we apply a minimum halo mass threshold of
$M_{\rm FoF}\geq M_{\rm min}=3.2\times10^{13}\,h^{-1}M_\odot$.
In addition, two halo mass threshold variation samples, denoted as Mmin\_p and Mmin\_m (corresponding to $M_{\rm min}^{+}$ and $M_{\rm min}^{-}$, respectively), are constructed from the 500 reserved fiducial realizations by adopting
$M_{\rm min}^{+}=3.3\times10^{13}\,h^{-1}M_\odot$
and
$M_{\rm min}^{-}=3.1\times10^{13}\,h^{-1}M_\odot$.
These samples are used to investigate the impact of halo selection on the resulting observables.

Since galaxy surveys measure the large-scale structure in redshift space, we convert the halo positions from real space to redshift space by including the contribution from halo peculiar velocities \cite{Scoccimarro_2004}. Under the distant-observer approximation, the redshift-space position of each halo is given by
\begin{equation}
\bm{s}
=
\bm{x}
+
\frac{\bm{v}\cdot\bm{n}}
{a(z)H(z)}
\bm{n},
\end{equation}
where $\bm{x}$ and $\bm{s}$ denote the halo positions in real and redshift space, respectively, $\bm{v}$ is the halo peculiar velocity, $\bm{n}$ is the line-of-sight direction, $a(z)=1/(1+z)$ is the scale factor, and $H(z)$ is the Hubble parameter. In this work, we adopt the plane-parallel approximation \cite{Castorina_2018} and set the line-of-sight direction along the $z$-axis.

After converting the halo positions from real space to redshift space, the halo positions are deposited onto a $384^3$ Cartesian grid using the Piecewise Cubic Spline (PCS) density assignment scheme \cite{Sefusatti_2016}, implemented with the Pylians3 library \cite{Pylians}.
Each halo is assigned equal weight in this procedure, resulting in a gridded redshift-space halo number density field. The PCS scheme is adopted to reduce aliasing effects and improve the accuracy of the density-field reconstruction over the wavenumber range considered in this work.

\section{Fisher formalism}\label{sec:fisher}

To quantitatively evaluate the cosmological information contained in the redshift-space monopole power spectra of halo density fields, including both the original halo power spectrum ($P_{\rm h,0}$) and the ZI power spectra ($P_{\rm ZI-\eta,0}$), we adopt the Fisher information formalism \cite{Tegmark_1997} to forecast cosmological parameter constraints. Assuming a Gaussian likelihood \cite{Upham_2021} for the data vector (see \ref{appendix} for a validation of this approximation), the Fisher matrix is given by
\begin{equation}
F_{ij}
=
\left(
\frac{\partial \boldsymbol{d}}
{\partial \theta_i}
\right)^{\rm T}
\bm{C}^{-1}
\left(
\frac{\partial \boldsymbol{d}}
{\partial \theta_j}
\right).
\end{equation}
where $\boldsymbol{d}$ denotes the data vector, $\boldsymbol{\theta}$ represents the set of cosmological parameters to be constrained, and $\bm{C}$ is the covariance matrix of the data vector. 

\figderivativescomparison

The covariance matrix is estimated from simulation samples and assumed to be independent of cosmological parameters,
\begin{equation}
\bm{C}
=
\left\langle
\left(\boldsymbol{d}-\left\langle\boldsymbol{d}\right\rangle\right)^{\rm T}
\left(\boldsymbol{d}-\left\langle\boldsymbol{d}\right\rangle\right)
\right\rangle.
\end{equation}
Figure~\ref{fig:correlation} compares the correlation matrix $c(k,k')=C_{kk'}/\sqrt{C_{kk}C_{k'k'}}$ of the combined data vector $\bm{d}$ before and after applying the noise correction to the ZI power spectra. Before the correction, the covariance matrix exhibits noticeable off-diagonal correlations, particularly toward higher wavenumbers, both within the ZI power spectra and between different components of the combined data vector. These correlations reflect the statistical dependencies induced by nonlinear structure formation and, for the ZI power spectra, the additional contribution associated with the discreteness of the halo field. After applying the proposed noise correction to the ZI power spectra, the off-diagonal correlations involving the ZI statistics are substantially reduced, particularly at high wavenumbers. This behavior suggests that the correction suppresses the additional covariance induced by halo discreteness while preserving the intrinsic power-spectrum signal, thereby providing a cleaner covariance structure for subsequent cosmological parameter inference.

The derivatives of the redshift-space monopole power spectrum with respect to cosmological parameters are estimated using the finite-difference method,
\begin{equation}
\frac{\partial \boldsymbol{d}}
{\partial\theta_i}
=
\frac{
\boldsymbol{d}(\theta_i+\Delta\theta_i)
-
\boldsymbol{d}(\theta_i-\Delta\theta_i)
}
{2\Delta\theta_i},
\end{equation}
where $\Delta\theta_i$ denotes the perturbation step size of the parameter $\theta_i$.
For the neutrino mass parameter $M_\nu$, we adopt the forward finite-difference formula
\begin{equation}
\frac{\partial \boldsymbol{d}}
{\partial M_\nu}
=
\frac{\boldsymbol{d}(4\Delta M_\nu)
-12\boldsymbol{d}(2\Delta M_\nu)
+32\boldsymbol{d}(\Delta M_\nu)
-21\boldsymbol{d}(M_\nu=0)}
{12\Delta M_\nu}
,
\end{equation}
where $\Delta M_\nu=0.1\,{\rm eV}$. This fourth-order finite-difference scheme utilizes the available simulations with $M_\nu=0.0$, $0.1$, $0.2$, and $0.4\,{\rm eV}$ to obtain a more accurate response of the monopole power spectrum to the neutrino mass.

Figure~\ref{fig:derivatives_comparison} compares the logarithmic derivatives $\partial\ln P/\partial\theta$ before (panel a) and after (panel b) the noise correction. For most parameters, the derivatives show similar overall trends on large scales, indicating that the noise correction generally preserves the large-scale parameter responses. A notable exception is $M_\nu$, for which the correction produces a pronounced change in the derivative even on relatively large scales. Toward smaller scales, the differences between the corrected and uncorrected derivatives become more evident, with changes in both their amplitudes and scale dependence depending on the parameter and statistic considered. For several parameters, including $h$, $n_{\rm s}$, $\Omega_{\rm m}$, and $\Omega_{\rm b}$, the correction mainly modifies the detailed small-scale behavior while leaving the overall response broadly intact. Importantly, the correction does not lead to a uniform enhancement of the derivative amplitudes across parameters and scales. Therefore, the improved parameter constraints obtained from the noise-corrected ZI power spectra can't be attributed simply to larger parameter derivatives. Instead, the resulting Fisher information depends on the combined effects of the modified parameter responses and the covariance structure of the ZI power spectra.

According to the Cram\'{e}r--Rao bound, we use the inverse Fisher matrix as the parameter covariance matrix and estimate the marginalized parameter uncertainties as
$
\sigma(\theta_i)
=
\sqrt{(F^{-1})_{ii}}.
$ In all the presented results, unless stated otherwise, the numerical estimation of derivatives and covariances has been made with $N_{\rm deriv}$ = 500 and $N_{\rm fid}$ = 14500 realizations, respectively. In \ref{appendix}, we examine the dependence of the results on $N_{\rm fid}$ and $N_{\rm deriv}$, and assess their numerical convergence.

\section{Results and discussions}\label{sec:results}

\tabnoisecorrected

\figcornerII

\subsection{The ZI transform before and after noise correction}

To evaluate the improvement of the ZI transform in extracting cosmological information and to validate the effectiveness of the proposed noise correction method, we perform a Fisher analysis by comparing the Fisher matrix constraints obtained from the halo power spectrum $P_{\rm h,0}$, individual ZI-$\eta$ ($\eta=1,10,+\infty$) power spectra, and their combinations. For all statistics, we adopt the same maximum wavenumber, $k_{\max}=0.5\, h\,{\rm Mpc}^{-1}$, to ensure a consistent comparison. Table~\ref{tab:noise_corrected} presents the marginalized $1\sigma$ Fisher constraints derived from $P_{\rm h,0}$, $P_{\rm ZI-1,0}$, $P_{\rm ZI-10,0}$, $P_{\rm log,0}$, and the combined power spectra
\begin{equation}
 P_{\rm ZI,0}=P_{\rm ZI-1,0}\oplus P_{\rm ZI-10,0}\oplus P_{\rm log,0}.
\end{equation}

Compared with the conventional halo power spectrum $P_{\rm h,0}$, the ZI-$\eta$ power spectra provide tighter constraints on several cosmological parameters. This improvement arises because the nonlinear ZI transform redistributes information contained in the higher-order $N$-point correlation functions of the halo density field into the two-point statistics of the transformed field, thereby allowing part of the information that is not captured by the conventional halo power spectrum to be accessed through the ZI power spectrum. In particular, the combined statistic $P_{\rm ZI,0}$ achieves significant improvements for parameters including $M_\nu$, $h$, $n_{\rm s}$, and the PNG parameters $f_{\rm NL}^{\rm equil}$ and $f_{\rm NL}^{\rm ortho}$. Without  noise correction, $P_{\rm ZI,0}$ improves the constraints on $M_\nu$, $h$, $n_{\rm s}$, $f_{\rm NL}^{\rm equil}$, and $f_{\rm NL}^{\rm ortho}$ by factors of approximately $2.8$, $2.7$, $4.7$, $4.9$, and $5.7$, respectively, compared with $P_{\rm h,0}$. Due to the presence of scale-dependent bias, the halo power spectrum can provide relatively tight constraints on local PNG, but is less effective in constraining the other types of PNG. In contrast, the ZI power spectra not only yield tighter constraints on local PNG, but also substantially improve the constraints on the equilateral and orthogonal PNG models.

The comparison of ZI-$\eta$ power spectra, however, reveals that this gain is not uniform across different parameters. While individual ZI-$\eta$ power spectra improve the constraints on parameters such as $h$, $n_{\rm s}$, $f_{\rm NL}^{\rm equil}$, and $f_{\rm NL}^{\rm ortho}$, they provide only limited improvements for some other parameters. For example, the constraints on $f_{\rm NL}^{\rm local}$ and $\Omega_{\rm m}$ are only marginally improved compared with $P_{\rm h,0}$, whereas parameters such as $\sigma_8$ and $M_{\rm min}$ do not necessarily benefit from ZI-$\eta$ power spectra. This indicates that different cosmological parameters couple differently to the nonlinear information extracted by the ZI transform.

A comparison between the constraints before and after noise correction further demonstrates the effectiveness of the proposed correction method. For individual ZI-$\eta$ power spectra, the correction leads to substantial improvements for most cosmological parameters. Taking $P_{\rm ZI-1,0}$ as an example, the noise correction improves the constraints on $M_\nu$, $h$, $n_{\rm s}$, $f_{\rm NL}^{\rm equil}$, and $f_{\rm NL}^{\rm ortho}$ by factors of approximately $3.0$, $1.9$, $1.7$, $2.0$, and $1.6$, respectively. Similar improvements are also observed for $P_{\rm ZI-10,0}$ and $P_{\rm log,0}$. This indicates that ZI-$\eta$ power spectra contain a non-negligible  noise contribution, which degrades their ability to extract cosmological information. By reducing this additional noise component, the proposed correction method allows the nonlinear information encoded in the ZI-$\eta$ power spectra to be utilized more efficiently. The significant improvement on $M_\nu$ is particularly worth noting, as it suggests that the ZI transform and the proposed shot noise correction are effective in recovering neutrino-induced signatures that are largely washed out in the halo power spectrum. However, the impact of the  noise correction is relatively modest for parameters such as $f_{\rm NL}^{\rm local}$ and $\Omega_{\rm m}$, for which the corresponding improvements in the parameter constraints are much smaller than those observed for the other parameters.

For the combined statistic $P_{\rm ZI,0}$,  noise correction also improves the parameter constraints, although the relative gain is smaller than that obtained for individual ZI-$\eta$ power spectra. This is expected because the multi-$\eta$ combination already incorporates complementary information from different nonlinear transforms and partially suppresses the impact of  noise contamination. Consequently, the additional gain from  noise correction is reduced compared with the individual $\eta$ parameter cases. After  noise correction, $P_{\rm ZI,0}$ provides substantial improvements for parameters such as $M_\nu$, $h$, $n_{\rm s}$, $M_{\rm min}$, $f_{\rm NL}^{\rm equil}$, and $f_{\rm NL}^{\rm ortho}$, with the constraints improved by factors of approximately $5.2$, $4.3$, $5.8$, $2.3$, $5.6$, and $5.5$, respectively, compared with $P_{\rm h,0}$. These results confirm that the proposed  noise correction method effectively improves the information extraction capability of ZI power spectra. Meanwhile, the joint analysis of multiple ZI-$\eta$ statistics provides a robust constraint by combining complementary information and reducing the sensitivity to  noise contamination.

 Moreover, combining the conventional halo power spectrum with the ZI power spectra, $P_{\rm h,0}\oplus P_{\rm ZI,0}$, further improves the constraints compared with $P_{\rm ZI,0}$ alone. This demonstrates that the halo power spectrum and ZI power spectra contain complementary information, and the ZI transform should be regarded as a complementary statistic rather than a replacement for the conventional power spectrum.

\subsection{Comparison between ZI and marked power spectra}

\figcornerIII

To benchmark the information-extraction capability of the ZI transform against an established nonlinear statistic, we use the marked power spectrum as a reference \cite{Massara_2023,Cowell_2024,massara_2024,ebina2026}. Specifically, before calculating the power spectrum, the density field is first smoothed on a characteristic scale $R$ to obtain the smoothed density contrast $\delta_R(\bm{x})$. The marked function \cite{White_2016,Philcox_2020} is then defined as
\begin{equation}
m(\bm{x})
=
\left(
\frac{1+\delta_s}
{1+\delta_s+\delta_R(\bm{x})}
\right)^p ,
\end{equation}
where $p$ controls the strength of the nonlinear weighting, while $\delta_s$ regulates the sensitivity of the marked function to the local density environment. Using this marked function, the marked density field is defined as
\begin{equation}
\delta_M(\bm{x})
=
\frac{m(\bm{x})[1+\delta(\bm{x})]}
{\bar m}
-1,
\end{equation}
where $\bar m$ denotes the mean value of the mark. The marked power spectrum is subsequently calculated from $\delta_M(\bm{x})$.

We select a representative set of marked parameters and construct four marked power spectra, denoted as $M_1$, $M_2$, $M_3$, and $M_4$. Each marked statistic is determined by three parameters: the smoothing scale $R$, the nonlinear weighting parameter $p$, and the density threshold $\delta_s$. In this work, we fix $p=1$ and construct different marked statistics by varying $R$ and $\delta_s$. Specifically, the four marked power spectra adopt the following parameter configurations \cite{Massara_2023}:
\[
\begin{aligned}
&M_1:(30\,h^{-1}{\rm Mpc},0.1),\quad
M_2:(25\,h^{-1}{\rm Mpc},0.25),\\
&M_3:(20\,h^{-1}{\rm Mpc},0.5),\quad
M_4:(30\,h^{-1}{\rm Mpc},0.5).
\end{aligned}
\]
We combine these marked power spectra as
\begin{equation}
\sum M_{i,0}
=
M_{1,0}
\oplus
M_{2,0}
\oplus
M_{3,0}
\oplus
M_{4,0},
\end{equation}
obtaining a joint marked statistic that incorporates information from different environmental scales.

\tabfishercompareall

The theoretical modeling of noise in marked statistics on nonlinear scales remains highly uncertain \cite{Massara_2023,massara_2024}. Therefore, we do not apply  noise corrections to the marked power spectra in the following comparison. For consistency, the ZI power spectra considered in this analysis are also not corrected for  noise. This allows us to compare the parameter constraining power of the ZI transform and marked power spectra under the same treatment of  noise.

Figure~\ref{fig:corner_2} presents the parameter constraints obtained from the halo power spectrum, the ZI power spectrum, and the marked power spectra. For several cosmological parameters, the ZI power spectrum provides constraints comparable to, or even tighter than, those obtained from the marked power spectrum. In particular, compared with the marked statistic $\sum M_{i,0}$, the ZI power spectrum provides stronger constraints on parameters such as $f^{\rm local}_{\rm NL}$, $M_\nu$, $n_{\rm s}$, and $\Omega_{\rm m}$. 

As shown in Figure~\ref{fig:corner_3} and Table~\ref{tab:fisher_compare_all}, jointly analyzing the halo power spectrum with either the ZI power spectrum or the marked power spectrum further improves parameter constraints. The comparison between the ZI and marked statistics further reveals that their information extraction capabilities depend on the cosmological parameters under consideration. $P_{\rm h,0} \oplus \sum M_{i,0}$ provides stronger constraints on two PNG parameters, $f_{\rm NL}^{\rm equil}$ and $f_{\rm NL}^{\rm ortho}$. Relative to $P_{\rm h,0}$ alone, the constraints are improved by factors of approximately $6.3$ and $7.3$, respectively, exceeding the corresponding gains obtained with $P_{\rm h,0} \oplus P_{\rm ZI,0}$. 

In contrast, $P_{\rm h,0} \oplus P_{\rm ZI,0}$ provides tighter constraints for many other parameters, with particularly notable improvements for $f_{\rm NL}^{\rm local}$, $M_\nu$, $n_s$, and $M_\text{min}$, compared with $P_{\rm h,0} \oplus \sum M_{i,0}$. While the marked power spectrum enhances signals from specific density environments through density-dependent weighting, the ZI transform acts on the full density field through a nonlinear mapping, enabling it to more effectively capture the neutrino-induced effects in underdense regions. In particular, the scale-dependent signatures of neutrino free-streaming are better preserved in the ZI power spectrum than in the marked spectrum.

The complementarity between the ZI and marked statistics is further demonstrated by their joint analysis. Combining the halo power spectrum with both nonlinear statistics, $P_{\rm h,0}\oplus P_{\rm ZI,0}\oplus \sum M_{i,0}$, achieves the strongest constraints among all combinations. Compared with the halo power spectrum alone, this combination improves the constraints on $f_{\rm NL}^{\rm local}$, $f_{\rm NL}^{\rm equil}$, $f_{\rm NL}^{\rm ortho}$, $M_\nu$, $h$, $n_{\rm s}$, $\Omega_{\rm m}$, $\Omega_{\rm b}$, $\sigma_8$, and $M_{\rm min}$ by factors of approximately $1.6$, $7.6$, $9.1$, $4.5$, $3.9$, $6.7$, $3.3$, $2.9$, $2.1$, and $4.6$, respectively. Therefore, combining the ZI and marked power spectra, together with the conventional halo power spectrum, can further exploit their complementary information and achieve tighter cosmological constraints from nonlinear large-scale structure.

\subsection{Cosmological information versus $k_\mathrm{max}$}

\figsigmak

To further quantify the information extraction efficiency of the ZI and marked power spectra across different nonlinear scales, and to assess the impact of the proposed noise correction on the ZI power spectra, we examine the dependence of the marginalized $1\sigma$ parameter constraints and signal-to-noise ratio (SNR) on the maximum wavenumber $k_{\max}$. Compared with the Fisher analysis performed at a fixed $k_{\max}$ above, this analysis provides a more direct view of how the constraining power and statistical signal of different statistics evolve as progressively smaller-scale nonlinear modes are included.

Figure~\ref{fig:sigma_k} shows the marginalized $1\sigma$ parameter constraints obtained from the different statistics as a function of $k_{\max}$. In general, the parameter constraints become progressively tighter as $k_{\max}$ increases, indicating that incorporating smaller-scale nonlinear modes provides additional cosmological information. For the ZI power spectrum, the noise-corrected results yield substantially tighter constraints than the corresponding uncorrected results throughout the entire $k_{\max}$ range considered, with the improvement being particularly pronounced for parameters such as $M_\nu$, $h$, and $\Omega_{\rm b}$. For parameters such as $f_{\rm NL}^{\rm local}$, $n_{\rm s}$, and $\Omega_{\rm m}$, the uncorrected ZI power spectrum already provides tighter constraints than the marked power spectrum over the entire range of $k_{\max}$ considered. The constraining advantage of the ZI power spectrum for these parameters is not specific to a particular choice of $k_{\max}$, but remains evident across the nonlinear scales explored here.

Over the entire range of $k_{\max}$ considered, the combination $P_{\rm h,0}\oplus P_{\rm ZI,0}$ consistently provides tighter constraints than either $P_{\rm h,0}$ or $P_{\rm ZI,0}$ alone. This indicates that, throughout the entire $k_{\max}$ range considered, the information contained in the ZI power spectrum does not fully encompass that of the conventional halo power spectrum, highlighting the complementary nature of the two statistics.

\figsnr

To further compare the effective utilization of statistical information contained in the different power spectra from a global perspective, we calculate their SNR as a function of $k_{\max}$. We define
\begin{equation} 
\mathrm{SNR}
=
\sqrt{
\left\langle \boldsymbol{d} \right\rangle^{\rm T}
\bm{C}^{-1}
\left\langle \boldsymbol{d} \right\rangle
},
\end{equation}
where $\left\langle \boldsymbol{d}\right\rangle$ denotes the mean data vector evaluated at the fiducial cosmology, and $\bm{C}$ is the corresponding covariance matrix. For a given covariance matrix, this quantity characterizes the overall statistical significance of the power-spectrum signal relative to its measurement uncertainty, and thus provides a global measure of the signal strength and statistical detectability of the corresponding statistic. It should be emphasized that the SNR is not itself a measure of the Fisher information for any specific cosmological parameter, but rather characterizes the overall detectable signal of the statistic. 

Figure~\ref{fig:snr} shows the SNR obtained from the different statistics as a function of $k_{\max}$. Before  noise correction, the ZI power spectrum exhibits the highest SNR among the statistics considered, and its SNR continues to increase with increasing $k_{\max}$. This indicates that the ZI power spectrum retains substantial statistically detectable signal as progressively smaller-scale nonlinear modes are included. However, the uncorrected ZI power spectrum also contains an additional contribution arising from the discreteness of the halo field, which we refer to as the discreteness contribution. Consequently, the difference between the SNRs before and after  noise correction characterizes the impact of this additional contribution on the overall statistical signal of the ZI power spectrum.

After applying the  noise correction, the SNR of the ZI power spectrum decreases, reflecting the removal of the discreteness contribution modeled in this work. Nevertheless, at $k_{\max}>0.3\,h\,{\rm Mpc}^{-1}$, the corrected ZI power spectrum still exhibits a higher SNR than the joint statistic $P_{\rm h,0}\oplus\sum_i M_{i,0}$. This is particularly noteworthy because the corrected ZI power spectrum no longer contains the additional discreteness contribution included in the uncorrected measurement. The persistence of a high SNR after correction therefore indicates that the enhanced statistical signal of the ZI power spectrum cannot be primarily attributed to halo discreteness. Instead, it is consistent with the ability of the nonlinear ZI transform to redistribute information associated with nonlinear structure formation into the two-point statistics of the transformed density field.

The dependence of the SNR on $k_{\max}$ further reveals differences in how the various statistics utilize increasingly small-scale modes. For the conventional halo power spectrum $P_{\rm h,0}$, the SNR initially increases rapidly but its growth becomes noticeably slower at $k_{\max}\sim0.25\,h\,{\rm Mpc}^{-1}$, after which it gradually approaches a plateau. This behavior indicates that, for the conventional two-point statistic, the additional statistically detectable signal gained from including progressively smaller-scale modes becomes increasingly limited. A similar trend is observed for the marked power spectra, whose SNR also gradually flattens toward larger $k_{\max}$. Thus, while the inclusion of nonlinear modes increases the overall statistical signal of both the conventional and marked power spectra, their SNRs show signs of saturation at the smaller scales considered here. In contrast, the SNR of the ZI power spectrum continues to increase with $k_{\max}$ over the full range of $k_{\max}=0.1$--$0.5\,h\,{\rm Mpc}^{-1}$, with no clear indication of saturation, and the behavior is observed both before and after noise correction. This is particularly meaningful for neutrino mass constraints, as the neutrino-induced effects are predominantly manifested on small-scale structures.

\section{Conclusions}\label{sec:conclusions}

In this work, we investigate the cosmological information content of the redshift-space monopole power spectrum of the ZI transformed halo density field. The ZI transform is a local nonlinear transform controlled by the parameter $\eta$, which changes the relative contributions of different density regimes and thereby alters how higher-order statistical information is encoded in the power spectrum of the transformed field. Using halo catalogs from the Quijote and Quijote-PNG simulations, we quantify the constraining power of ZI power spectra with Fisher-matrix analyses.

We consider $\eta\geq1$ to avoid the numerical instability caused by the strong amplification of extremely underdense regions for $\eta<1$. We adopt $\eta=1$, $10$, and $+\infty$, and denote their combined power spectra as $P_{\rm ZI,0}$. Even without noise correction, $P_{\rm ZI,0}$ substantially improves the constraints relative to the conventional halo power spectrum $P_{\rm h,0}$. These results demonstrate that the ZI transform efficiently recovers nonlinear information that is not fully captured by the conventional halo power spectrum.

We also compare the ZI power spectrum with the marked power spectrum. The two statistics exhibit different sensitivities to the cosmological and PNG parameters, indicating that they contain complementary information. In particular, the ZI power spectrum provides tighter constraints on parameters such as $f_{\rm NL}^{\rm local}$ and $M_\nu$, while its signal-to-noise ratio continues to increase more effectively with $k_{\max}$, with less pronounced saturation at small scales. In addition, unlike the marked power spectrum, whose performance depends on the choice of mark parameters, the ZI transform is controlled by a single parameter $\eta$, making it less dependent on detailed parameter choices. Combining $P_{\rm h,0}$, $P_{\rm ZI,0}$, and $\sum M_{i,0}$ further tightens the constraints for most parameters relative to $P_{\rm h,0}$ alone.

We further address the noise contamination of ZI power spectra by introducing a scale-dependent noise model, $A=A(k\vert{}\eta)$. We estimate $A(k\vert{}\eta)$ using halo catalogs with different number densities generated by random downsampling. After correction, the power spectra obtained from catalogs with different sampling densities agree well, supporting the consistency of the proposed model. The amplitude of $A(k\vert{}\eta)$ also varies with $\eta$, and its scale dependence shows a similar trend to that of the corresponding ZI power spectra. This provides motivation for developing a theoretical model of $A(k\vert{}\eta)$ in future work. The correction further improves the cosmological constraints.

Overall, our results show that ZI power spectra provide an efficient two-point statistic for extracting nonlinear cosmological information from halo fields, while offering complementary information to both the conventional and marked power spectra. The scale-dependent noise correction further enhances their constraining power. Since the halo number densities used here are substantially lower than those expected for future galaxy surveys, our forecasts may be conservative. Future work will extend this analysis to galaxy samples, higher redshift-space multipoles, and theoretical modeling of $A(k|\eta)$ using nonlinear perturbation theory or the effective field theory of large-scale structure. These developments will help establish the applicability of ZI statistics to forthcoming large-scale structure surveys.


\Acknowledgements{This work was supported by the Start-up Funding from Jilin University (NO.~419080525A84), the Young Scientists Fund (C) of the National Natural Science Foundation of China (NO.~12603010), the Chinese Academy of Sciences President’s International Fellowship Initiative (NO.~2026PVA00044), and a grant from the Jilin Province Science and Technology Development Plan Item (NO.~20250205078GH).}%

\InterestConflict{The authors declare that they have no conflict of interest.}


\bibliographystyle{scpma}
\bibliography{ref}{}

\begin{appendix}

\renewcommand{\thesection}{Appendix}

\section{}\label{appendix}

To assess the dependence of the Fisher analysis on the number of simulations, we separately investigate the impact of the number of fiducial simulations used to construct the covariance matrix, $N_{\rm fid}$, and the number of simulations used to compute the numerical derivatives, $N_{\rm deriv}$, on the parameter constraints. 

\figfiddevtest
\tabconvergence
\figgautest

Figure.~\ref{fig:fidtest} shows the marginalized errors of the cosmological parameters, $\sigma(\theta_i\,|\,N_{\rm fid})/\sigma(\theta_i\,|\,N_{\rm fid}=14500)$, as a function of the number of fiducial simulations, $N_{\rm fid}$, used to calculate the covariance matrix. We consider $P_{\rm h,0}$, $P_{\rm ZI,0}$, the noise-corrected $P_{\rm ZI,0}$, $\sum M_{i,0}$, and different combinations of these statistics. As $N_{\rm fid}$ increases, the parameter constraints from all statistics gradually converge, with their overall variations becoming significantly smaller. In particular, for $N_{\rm fid}\geq 10000$, the marginalized errors of the cosmological parameters are all within $1\%$. This indicates that, with the number of fiducial simulations adopted in this work, the impact of statistical noise in the covariance matrix on the final Fisher parameter constraints is under control.

Figure.~\ref{fig:devtest} shows the marginalized errors of the cosmological parameters, $\sigma(\theta_i,|,N_{\rm deriv})/\sigma_{\rm fit}(\theta_i,|,N_{\rm deriv}=5000)$, as a function of the number of simulations, $N_{\rm deriv}$, used to calculate the derivatives. Clearly, the parameter constraints have not yet converged for $N_{\rm deriv}\leq500$. To quantitatively describe the asymptotic convergence of the constraints with increasing $N_{\rm deriv}$, we fit the convergence curves with $\sigma(N_{\rm deriv})=a+b\left[1-\exp\left(-cN_{\rm deriv}\right)\right]$, where $a$, $b$, and $c$ are fitting parameters. This functional form is chosen because the estimated marginalized errors are expected to asymptotically converge to their true values as the number of simulations increases, where the true values correspond to those obtained when calculating the derivatives with an infinite number of simulations. The points in the figure represent the Fisher parameter constraints directly calculated at different $N_{\rm deriv}$, while the solid curves show the best-fit results. It can be seen that, for $P_{\rm ZI,0}$ and $P_{\rm h,0}\oplus P_{\rm ZI,0}$, the parameter constraints generally become stable at relatively small $N_{\rm deriv}$ compared with those obtained from $\sum M_{i,0}$ and $P_{\rm h,0}\oplus\sum M_{i,0}$. After noise correction, $P_{\rm ZI,0}$ and $P_{\rm h,0}\oplus P_{\rm ZI,0}$ require fewer derivative simulations, $N_{\rm deriv}$, to achieve converged constraints on the neutrino mass parameter $M_\nu$. In contrast, the conventional halo power spectrum $P_{\rm h,0}$ shows a more pronounced dependence on $N_{\rm deriv}$ for some parameters. Overall, $N_{\rm deriv}=5000$ provides a relatively stable estimate of the derivatives and is therefore adopted as the reference value in this convergence test.

Table~\ref{tab:convergence} presents the marginalized parameter constraints extrapolated to $N_{\rm deriv}=5000$. It should be emphasized that these results are not obtained by actually running 5000 independent simulations for calculating the derivatives, but by fitting the convergence curves measured from the 500 available simulations and then extrapolating the fitted functions to $N_{\rm deriv}=5000$. As can be seen from the table, except for $f_{\rm NL}^{\rm equil}$ and $f_{\rm NL}^{\rm ortho}$, $P_{\rm h,0}\oplus P_{\rm ZI,0}$ provides tighter constraints than $P_{\rm h,0}\oplus\sum M_{i,0}$ for all other parameters. In particular, the results from the noise-corrected $P_{\rm h,0}\oplus P_{\rm ZI,0}$ can further improve the constraints.

We next test whether the statistics adopted in this work can be reasonably approximated as multivariate Gaussian distributed. Following Friedrich et al. \cite{Friedrich_2021}, we compute the $\chi^2$ value for each data vector measured from the 14,500 fiducial simulations,
\begin{equation}
\chi^2
=
\left(
\boldsymbol{d} - \langle\boldsymbol{d}\rangle
\right)
\bm{C}^{-1}
\left(
\boldsymbol{d} - \langle\boldsymbol{d}\rangle
\right)^{\rm T},   
\end{equation}
where $\boldsymbol d$ denotes the data vector measured from the fiducial simulation, $\langle\boldsymbol{d}\rangle$ is the mean data vector over all fiducial simulations, and $\boldsymbol C$ is the covariance matrix estimated from these simulations.

As shown in Figure \ref{fig:gau_test}, we perform this test for $P_{\rm h,0}$, $P_{\rm ZI,0}$, the noise-corrected $P_{\rm ZI,0}$, $\sum M_{i,0}$, and their various combinations. For all statistics, the $\chi^2$ distributions obtained from the fiducial simulations show overall good agreement with those obtained from random realizations of a multivariate Gaussian distribution with the same mean and covariance, as well as with the corresponding theoretical $\chi^2$ distributions. This indicates that the multivariate Gaussian approximation provides a reasonable description of the statistical distributions of the power-spectrum data vectors considered in this work.

\end{appendix}

\end{multicols}

\end{document}

%% file: figures.tex
\newcommand{\figZI}{
\begin{figure}[H]
\centering
\includegraphics[scale=0.5]{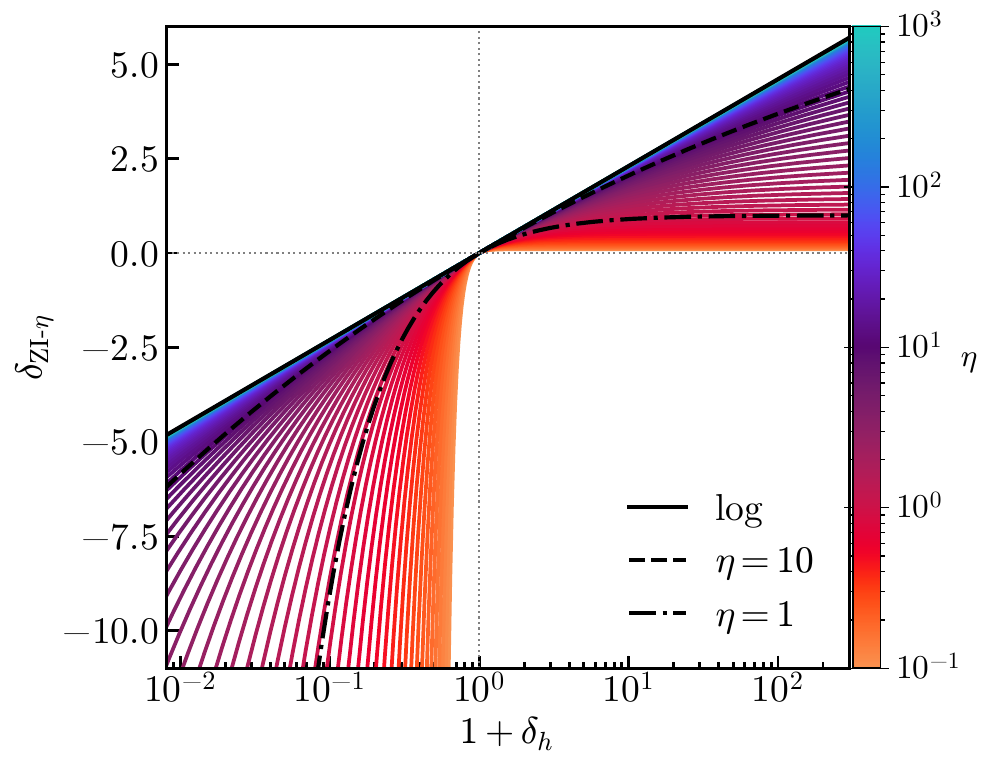}
\caption{The relation between the halo overdensity $1+\delta_{\rm h}$ and the transformed overdensity $\delta_{\rm ZI-\eta}$ for different values of $\eta$.} 
\label{fig:ZI}
\end{figure}
}

\newcommand{\fignoise}{
\begin{figure}[H]
\centering
\includegraphics[scale=0.5]{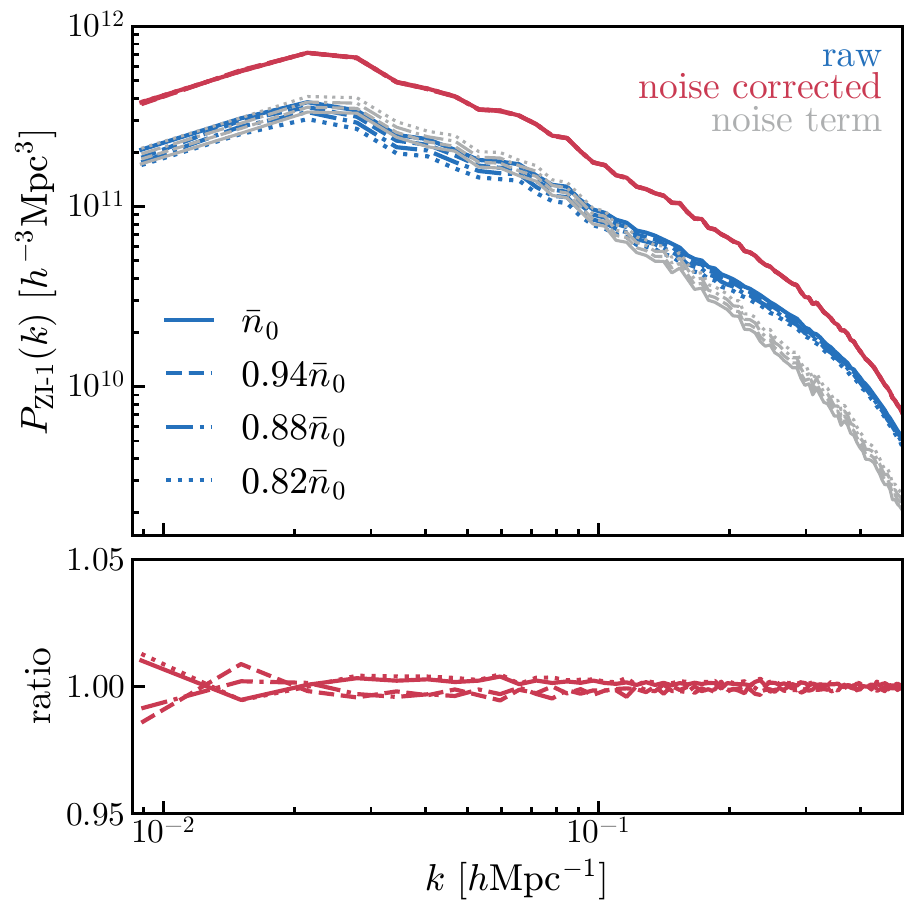}
\caption{Scale-dependent noise correction for the ZI power spectrum at $\eta=1$ using halo catalogs with different mean number densities. The upper panel shows the raw, noise contribution, and noise-corrected power spectra, while the lower panel shows the ratios of the corrected spectra to their mean. The close agreement among the corrected spectra demonstrates the effectiveness of the noise correction.}
\label{fig:noise}
\end{figure}
}

\newcommand{\figeta}{
\begin{figure}[H]
\centering
\includegraphics[scale=0.5]{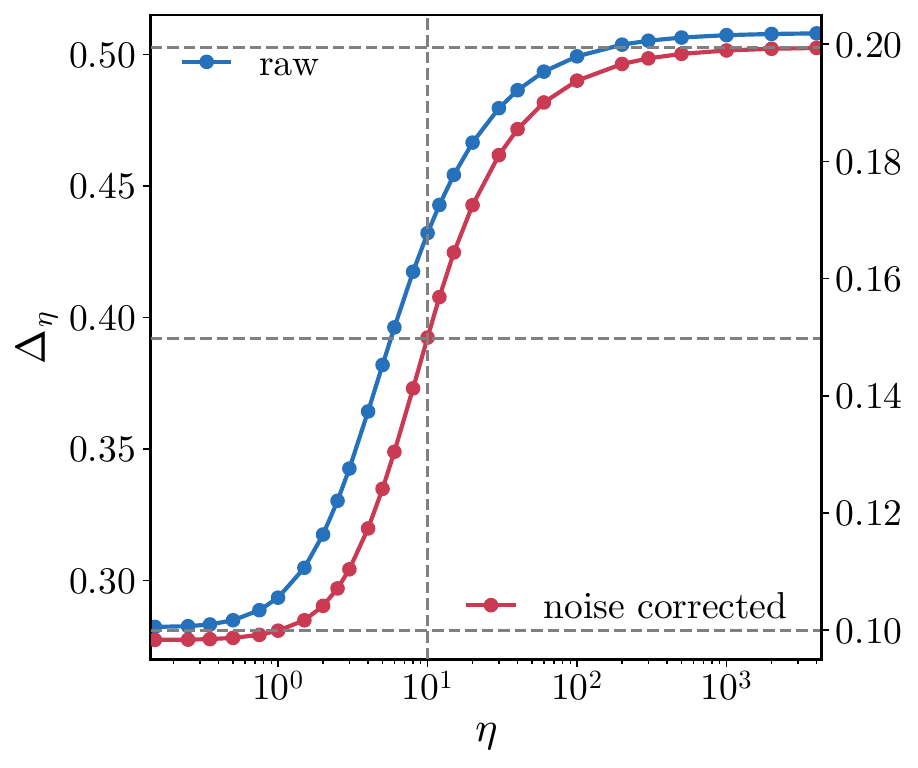}
\caption{The dependence of $\Delta_\eta$ on the ZI transform parameter $\eta$. The blue curve shows the result obtained from the raw ZI power spectra before noise correction and corresponds to the left vertical axis, while the red curve shows $\Delta_\eta$ calculated from the noise-corrected ZI power spectra and corresponds to the right vertical axis.} 
\label{fig:eta}
\end{figure}
}

\newcommand{\figcorrelation}{
\begin{figure*}[t]
    \centering
    \includegraphics[width=1.8\columnwidth]{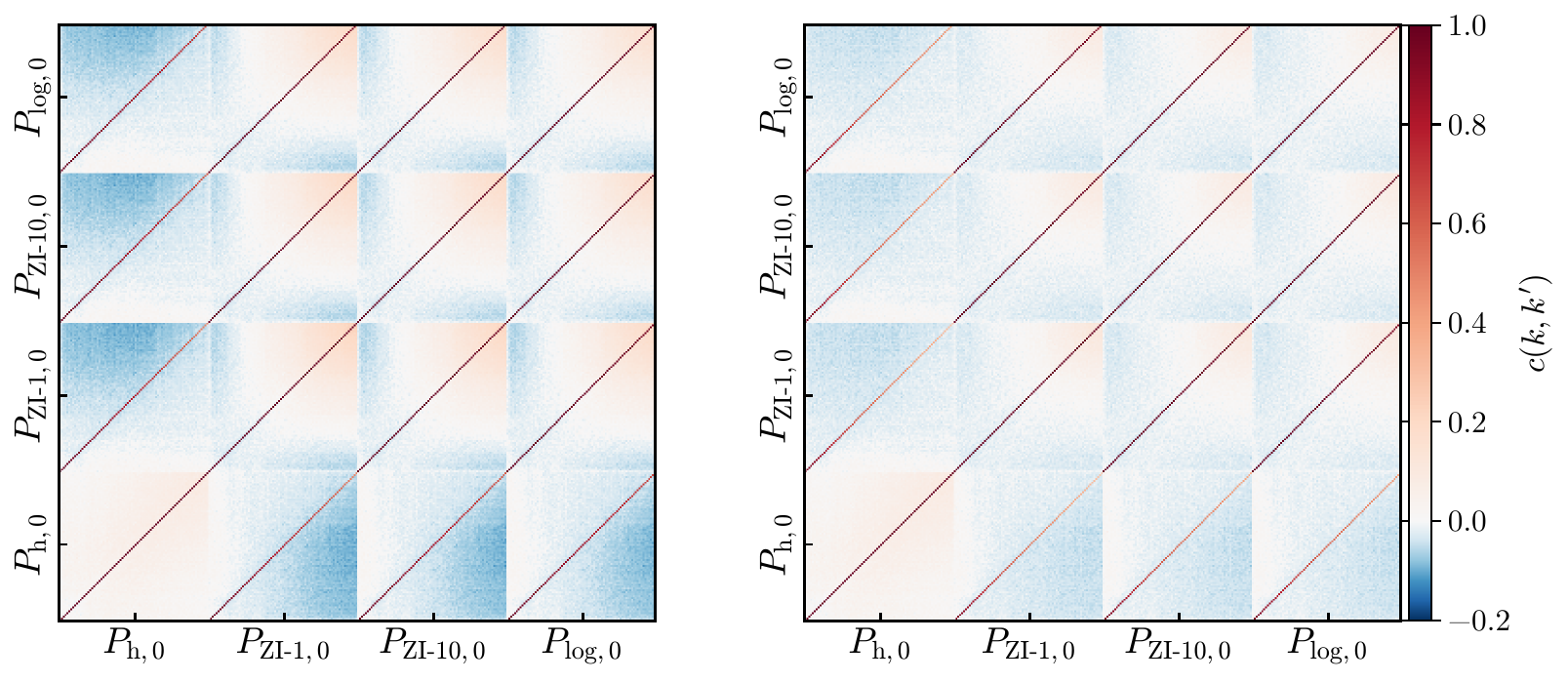}
    \caption{
    Comparison of the correlation matrices for the combined data vector
    $P_{\rm h,0}\oplus P_{\rm ZI-1,0}\oplus P_{\rm ZI-3,0}\oplus P_{\rm log,0}$
    before (left) and after (right) applying the noise correction to the ZI power spectra. The matrices are computed using Fourier
    modes up to $k_{\rm max}=0.5\,h\,{\rm Mpc}^{-1}$. Note that the halo power spectrum $P_{\rm h,0}$ in both panels has already been corrected by subtracting the Poisson shot-noise term $1/\bar n$.
    }
    \label{fig:correlation}
\end{figure*}
}

\newcommand{\figderivativescomparison}{
\begin{figure*}[t]
\centering

\includegraphics[width=0.95\textwidth]{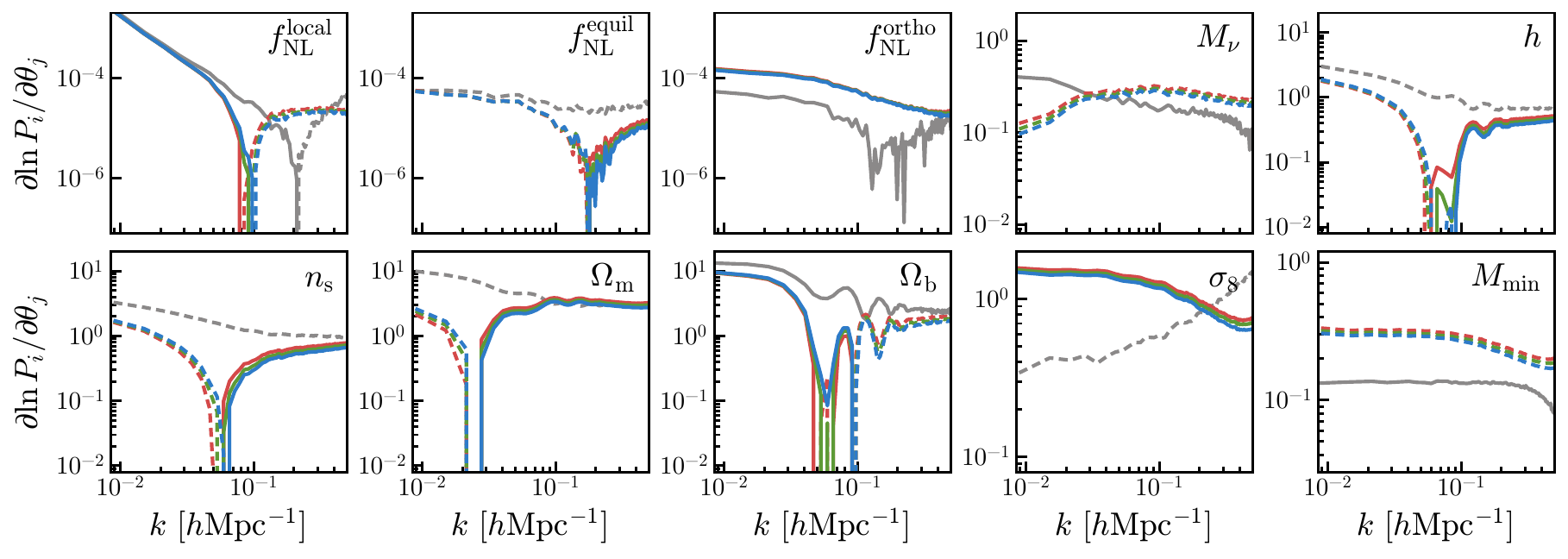}

\vspace{-2mm}
{\fontsize{8}{10}\selectfont (a)}

\vspace{3mm}

\includegraphics[width=0.95\textwidth]{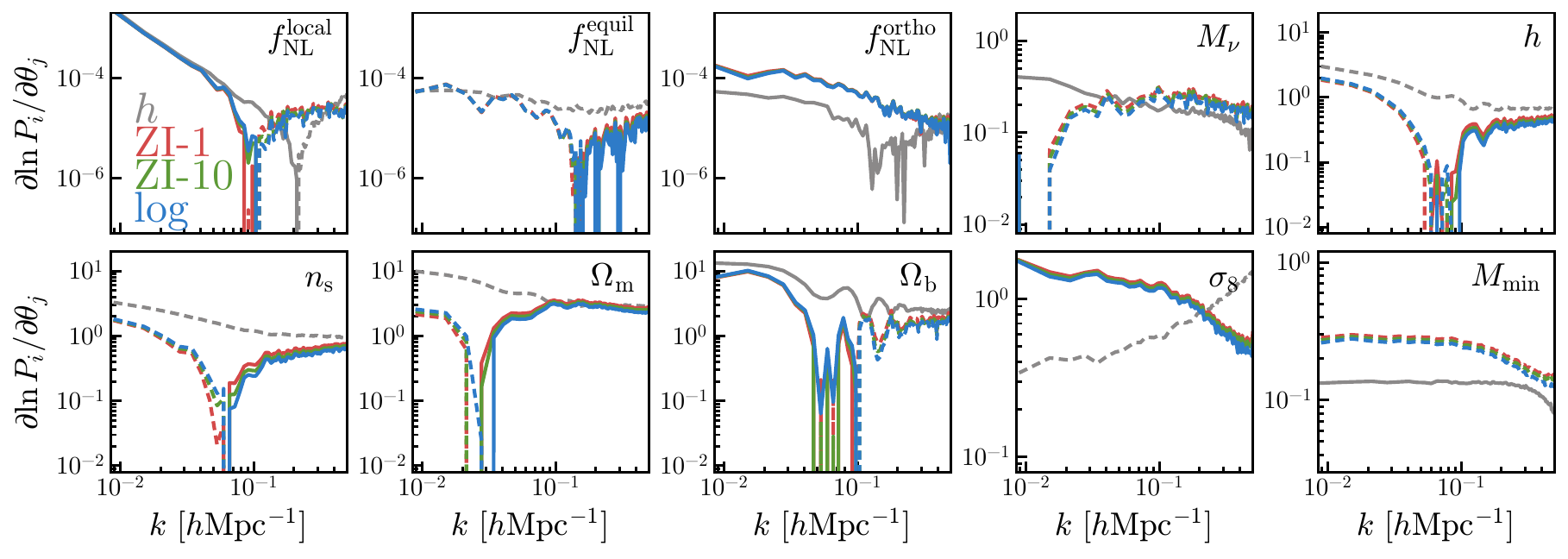}

\vspace{-2mm}
{\fontsize{8}{10}\selectfont (b)}

\caption{Derivatives of the logarithm of the power spectra with respect to the parameters, $\partial\ln P/\partial\theta$, before (panel a) and after (panel b) applying the proposed noise correction to the ZI power spectra. The different curves correspond to the halo power spectrum $P_{\rm h,0}$ and the three ZI-transformed power spectra, $P_{\rm ZI-1,0}$, $P_{\rm ZI-3,0}$, and $P_{\rm log,0}$.}
\label{fig:derivatives_comparison}

\end{figure*}
}

\newcommand{\figcornerII}{
\begin{figure*}[t]
    \centering
    \includegraphics[width=1.8\columnwidth]{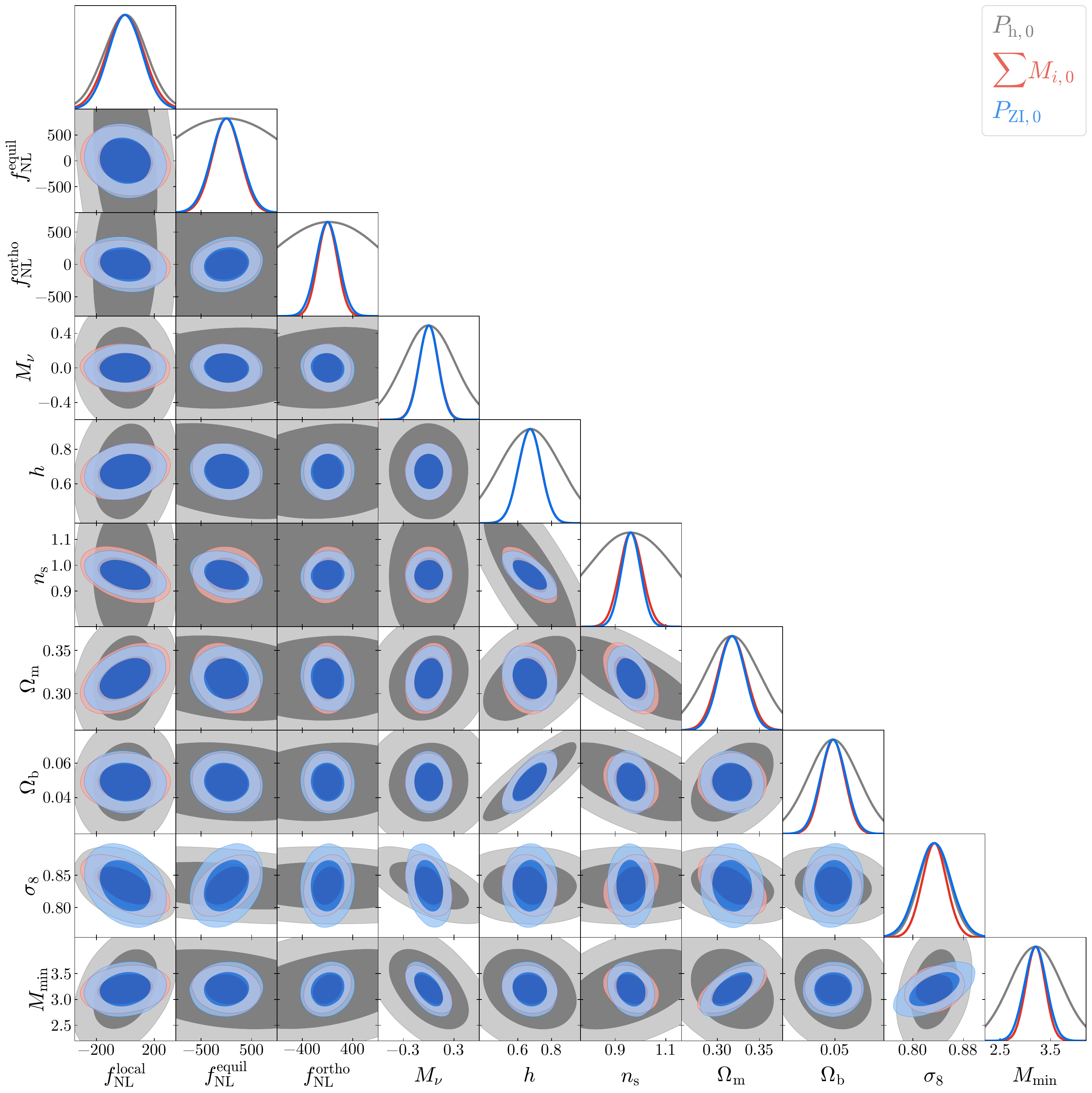}
    \caption{
    Fisher forecasted constraints on the parameters from different power spectra. 
    The gray, red, and blue contours correspond to $P_{\rm h,0}$, $\sum M_{i,0}$, and $P_{\rm{ZI},0}$, respectively. 
    }
    \label{fig:corner_2}
\end{figure*}
}

\newcommand{\figcornerIII}{
\begin{figure*}[t]
    \centering
    \includegraphics[width=1.8\columnwidth]{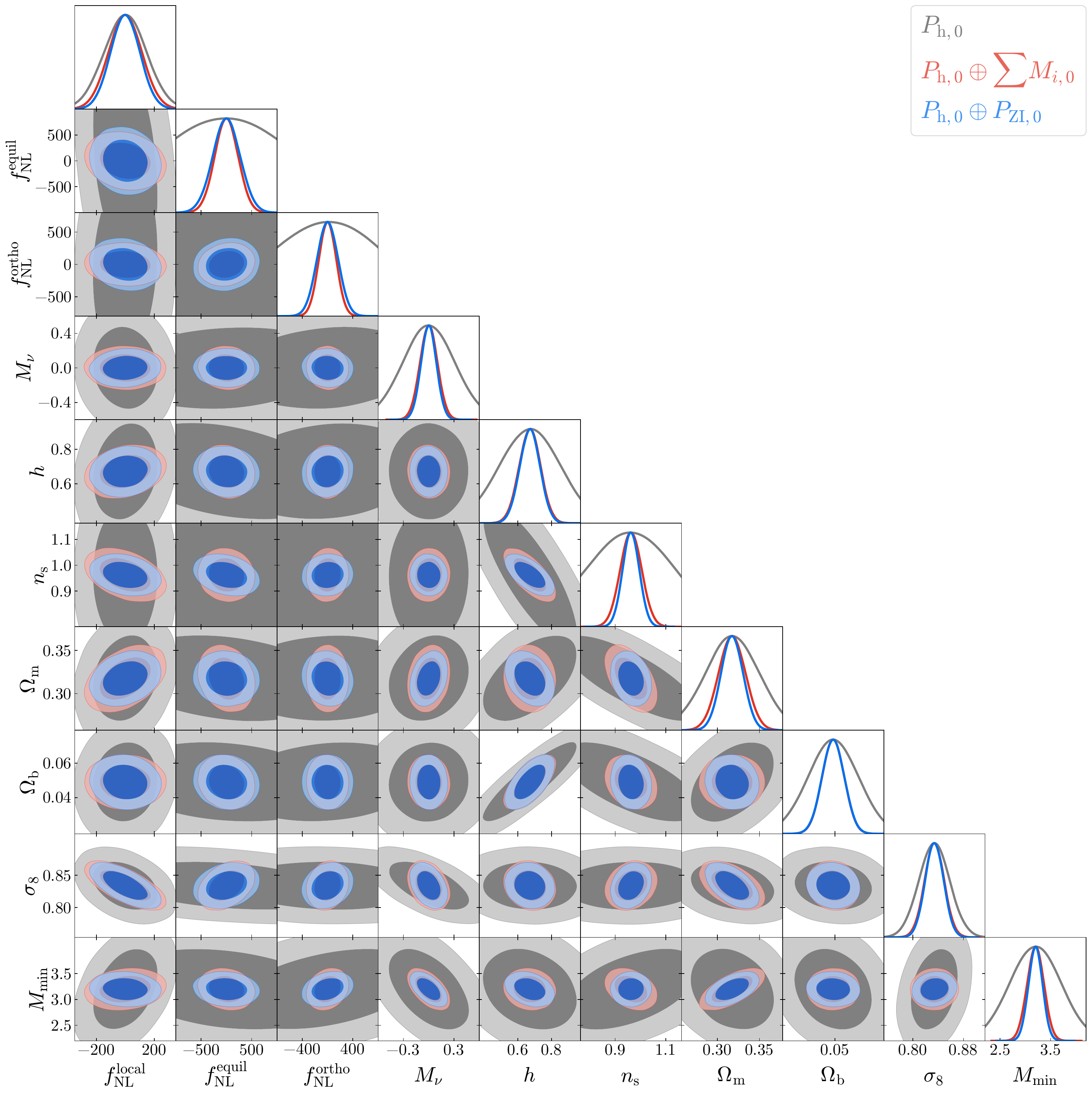}
    \caption{
    The same as Figure \ref{fig:corner_2}, but for $P_{\rm h,0}$ (gray), $P_{\rm h,0}\oplus\sum M_{i,0}$ (red), and $P_{\rm h,0}\oplus P_{\rm{ZI},0}$ (blue). 
    }
    \label{fig:corner_3}
\end{figure*}
}

\newcommand{\figsigmak}{
\begin{figure*}[t]
    \centering
    \includegraphics[width=2.0\columnwidth]{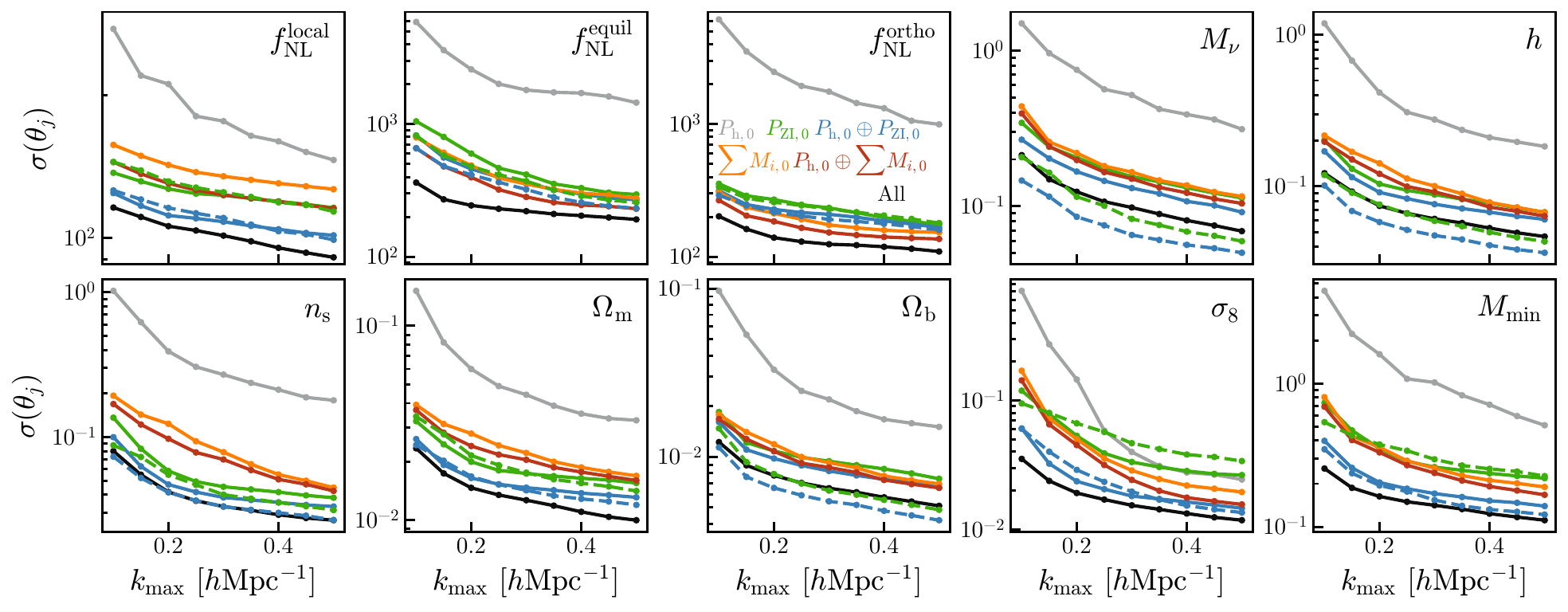}
    \caption{The marginalized $1\sigma$ parameter constraints obtained from the different statistics are shown as a function of $k_{\max}$. The solid lines denote the results without noise correction, while the dashed lines represent the results after applying the noise correction method proposed in this work. The statistic \textit{All} denotes the combination $P_{\rm h,0}\oplus P_{\rm{ZI},0}\oplus\sum M_{i,0}$.}
    \label{fig:sigma_k}
\end{figure*}
}

\newcommand{\figsnr}{
\begin{figure}[H]
\centering
\includegraphics[scale=0.5]{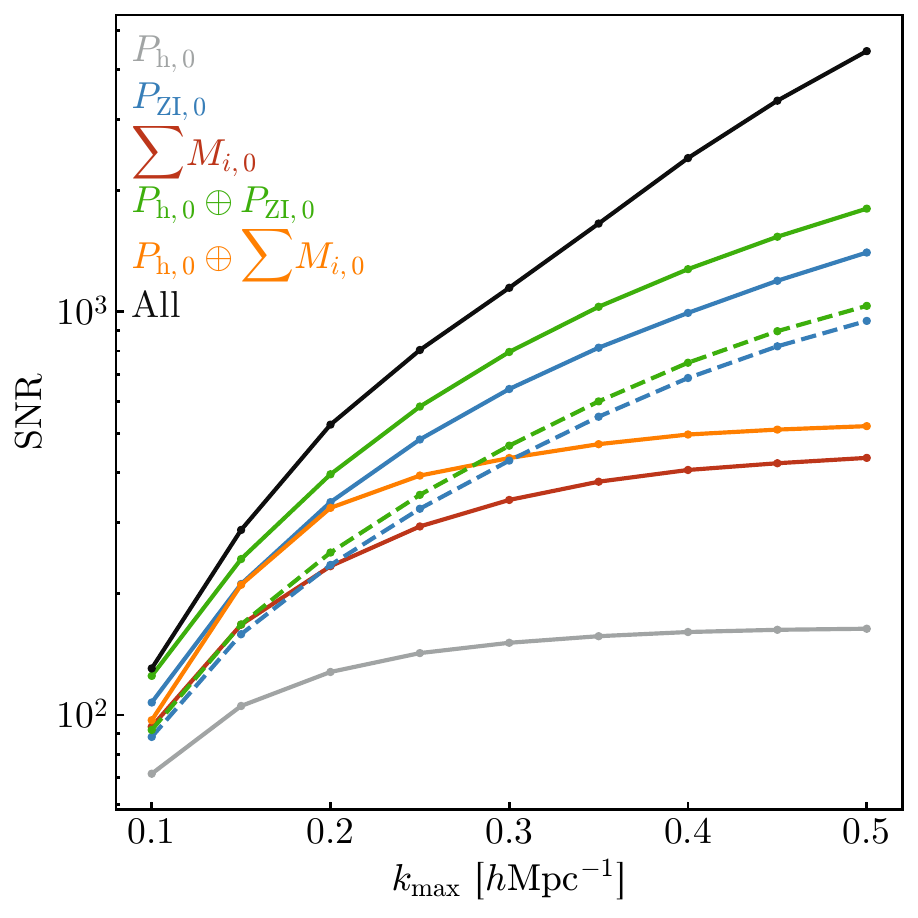}
\caption{Signal-to-noise ratios (SNRs) of the different statistics as a function of $k_{\max}$. The solid lines denote the results without noise correction, while the dashed lines represent the results after applying the noise correction method proposed in this work. The statistic \textit{All} denotes the combination $P_{\rm h,0}\oplus P_{\rm{ZI},0}\oplus\sum M_{i,0}$.} 
\label{fig:snr}
\end{figure}
}

\newcommand{\figfiddevtest}{
\begin{figure*}[!t]
    \centering

    \includegraphics[width=2.0\columnwidth]{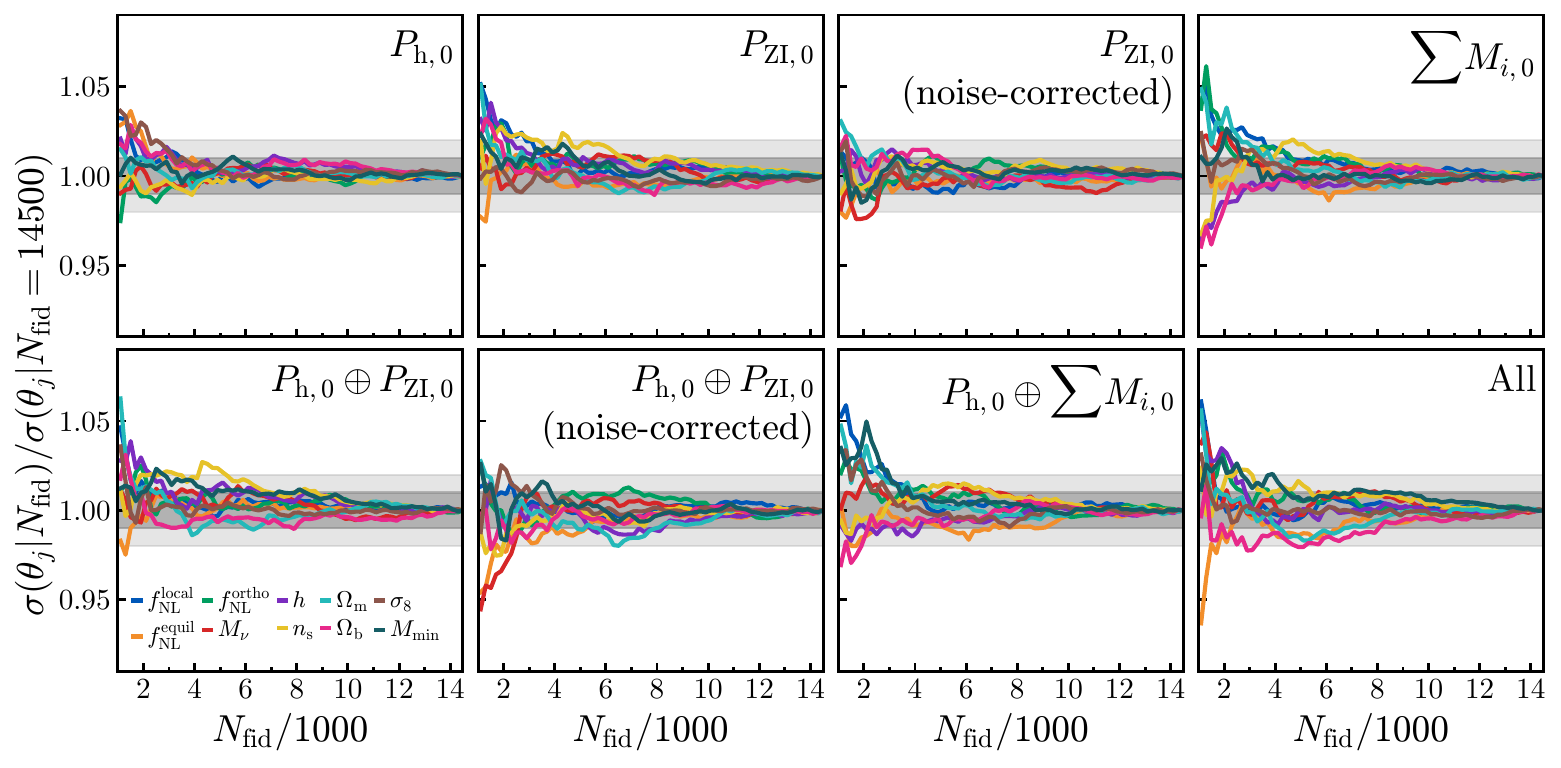}

    \caption{Convergence of the marginalized $1\sigma$ Fisher constraints as a function of the number of fiducial simulations, $N_{\rm fid}$.
    The constraints are normalized by the corresponding values obtained from $N_{\rm fid}=14500$ fiducial simulations.
    The horizontal dark- and light-gray bands indicate $\pm1\%$ and $\pm2\%$ deviations from the reference value, respectively.}
    \label{fig:fidtest}

    \vspace{0.5cm}

    \includegraphics[width=2.0\columnwidth]{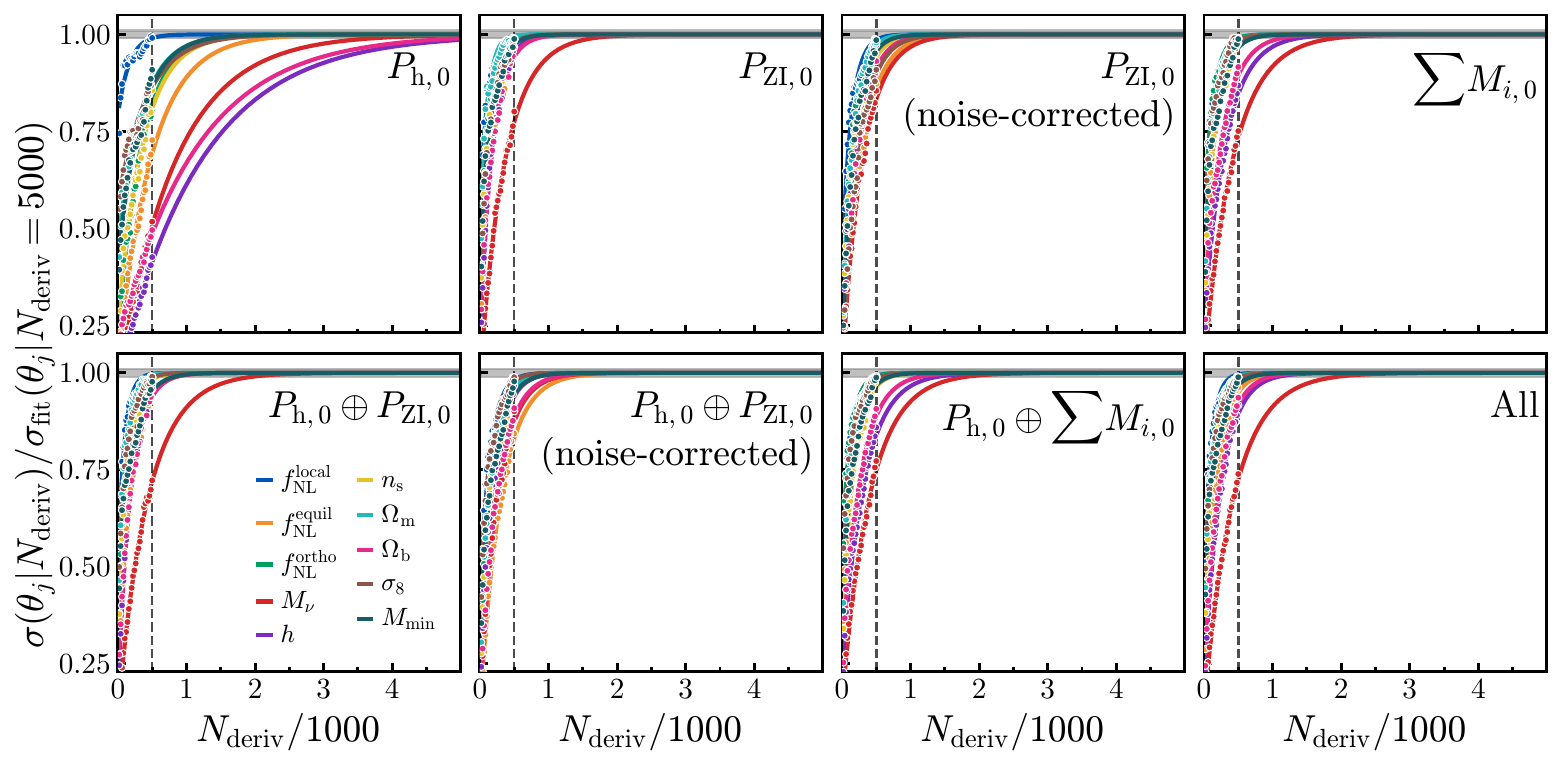}

    \caption{Convergence of the marginalized $1\sigma$ Fisher constraints as a function of the number of derivative simulations, $N_{\rm deriv}$.
    The colored points show the constraints obtained from different numbers of derivative simulations, while the solid curves show the corresponding convergence fits extrapolated to $N_{\rm deriv}=5000$.
    The constraints are normalized by the fitted value at $N_{\rm deriv}=5000$.
    The vertical dashed line marks $N_{\rm deriv}=500$, while the horizontal gray bands indicate a $\pm1\%$ deviation from the extrapolated asymptotic value.}
    \label{fig:devtest}

\end{figure*}
}

\newcommand{\figgautest}{
\begin{figure*}[t]
    \centering
    \includegraphics[width=2.0\columnwidth]{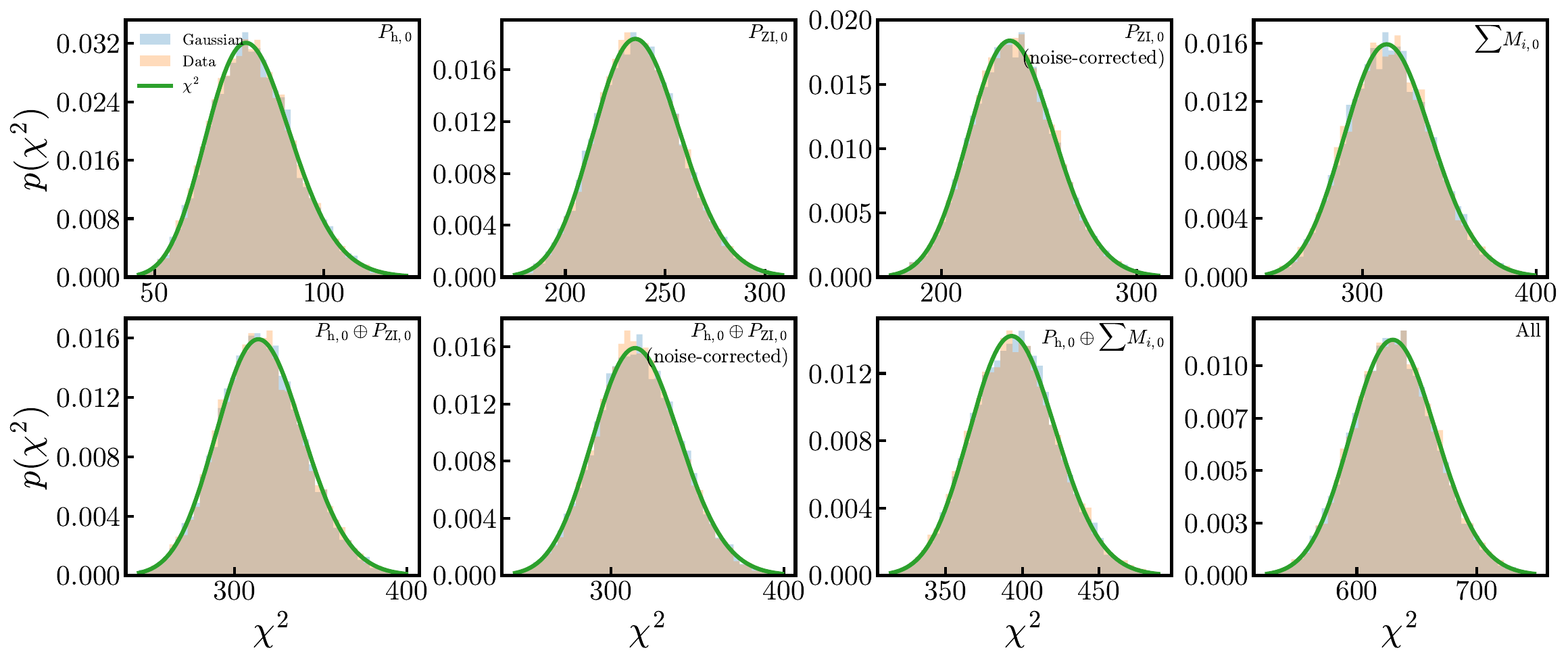}
    \caption{Gaussianity test for the eight power-spectrum data vectors. The distributions of $\chi^2$ measured from the 14,500 fiducial simulations (Data) are compared with those from multivariate Gaussian realizations (Gaussian), while the solid curves show the corresponding theoretical $\chi^2$ distributions. The panels show the different combinations of halo, ZI, and marked power spectra, including the noise-corrected cases.}
    \label{fig:gau_test}
\end{figure*}
}

%% file: tables.tex
\newcommand{\tabsimulations}{
\begin{table*}[t]
\centering
\caption{Summary of the simulation datasets and their parameter configurations.}
\label{tab:simulations}
\resizebox{\textwidth}{!}{%
\renewcommand{\arraystretch}{0.85}
\begin{tabular}{llccccccccccc}
\toprule
\addlinespace[3pt]
Set & ICs & $f_\text{NL}^\text{local}$ & $f_\text{NL}^\text{equil}$ & $f_\text{NL}^\text{ortho}$ & $M_\nu \, [\text{eV}]$ & $h$ & $n_{\rm s}$ & $\Omega_{\rm m}$ & $\Omega_{\rm b}$ & $\sigma_8$ & $M_\text{min} \, [10^{12}h^{-1}M_\odot]$ & Realizations \\
\addlinespace[3pt]
\midrule
\addlinespace[3pt]
fiducial & 2LPT & 0 & 0 & 0 & 0 & 0.6711 & 0.9624 & 0.3175 & 0.049 & 0.834 & 3.2 & 14,500 \\
LC\_p & 2LPTPNG & \textbf{+100} & 0 & 0 & 0 & 0.6711 & 0.9624 & 0.3175 & 0.049 & 0.834 & 3.2 & 500 \\
LC\_m & 2LPTPNG & \textbf{-100} & 0 & 0 & 0 & 0.6711 & 0.9624 & 0.3175 & 0.049 & 0.834 & 3.2 & 500 \\
EQ\_p & 2LPTPNG & 0 & \textbf{+100} & 0 & 0 & 0.6711 & 0.9624 & 0.3175 & 0.049 & 0.834 & 3.2 & 500 \\
EQ\_m & 2LPTPNG & 0 & \textbf{-100} & 0 & 0 & 0.6711 & 0.9624 & 0.3175 & 0.049 & 0.834 & 3.2 & 500 \\
OR\_LSS\_p & 2LPTPNG & 0 & 0 & \textbf{+100} & 0 & 0.6711 & 0.9624 & 0.3175 & 0.049 & 0.834 & 3.2 & 500 \\
OR\_LSS\_m & 2LPTPNG & 0 & 0 & \textbf{-100} & 0 & 0.6711 & 0.9624 & 0.3175 & 0.049 & 0.834 & 3.2 & 500 \\
fiducial\_ZA & ZA & 0 & 0 & 0 & 0 & 0.6711 & 0.9624 & 0.3175 & 0.049 & 0.834 & 3.2 & 500 \\
Mnu\_p & ZA & 0 & 0 & 0 & \textbf{0.1} & 0.6711 & 0.9624 & 0.3175 & 0.049 & 0.834 & 3.2 & 500 \\
Mnu\_pp & ZA & 0 & 0 & 0 & \textbf{0.2} & 0.6711 & 0.9624 & 0.3175 & 0.049 & 0.834 & 3.2 & 500 \\
Mnu\_ppp & ZA & 0 & 0 & 0 & \textbf{0.4} & 0.6711 & 0.9624 & 0.3175 & 0.049 & 0.834 & 3.2 & 500 \\
h\_p & 2LPT & 0 & 0 & 0 & 0 & \textbf{0.6911} & 0.9624 & 0.3175 & 0.049 & 0.834 & 3.2 & 500 \\
h\_m & 2LPT & 0 & 0 & 0 & 0 & \textbf{0.6511} & 0.9624 & 0.3175 & 0.049 & 0.834 & 3.2 & 500 \\
ns\_p & 2LPT & 0 & 0 & 0 & 0 & 0.6711 & \textbf{0.9824} & 0.3175 & 0.049 & 0.834 & 3.2 & 500 \\
ns\_m & 2LPT & 0 & 0 & 0 & 0 & 0.6711 & \textbf{0.9424} & 0.3175 & 0.049 & 0.834 & 3.2 & 500 \\
Om\_p & 2LPT & 0 & 0 & 0 & 0 & 0.6711 & 0.9624 & \textbf{0.3275} & 0.049 & 0.834 & 3.2 & 500 \\
Om\_m & 2LPT & 0 & 0 & 0 & 0 & 0.6711 & 0.9624 & \textbf{0.3075} & 0.049 & 0.834 & 3.2 & 500 \\
Ob2\_p & 2LPT & 0 & 0 & 0 & 0 & 0.6711 & 0.9624 & 0.3175 & \textbf{0.051} & 0.834 & 3.2 & 500 \\
Ob2\_m & 2LPT & 0 & 0 & 0 & 0 & 0.6711 & 0.9624 & 0.3175 & \textbf{0.047} & 0.834 & 3.2 & 500 \\
s8\_p & 2LPT & 0 & 0 & 0 & 0 & 0.6711 & 0.9624 & 0.3175 & 0.049 & \textbf{0.849} & 3.2 & 500 \\
s8\_m & 2LPT & 0 & 0 & 0 & 0 & 0.6711 & 0.9624 & 0.3175 & 0.049 & \textbf{0.819} & 3.2 & 500 \\
Mmin\_p & 2LPT & 0 & 0 & 0 & 0 & 0.6711 & 0.9624 & 0.3175 & 0.049 & 0.834 & \textbf{3.3} & 500$^*$ \\
Mmin\_m & 2LPT & 0 & 0 & 0 & 0 & 0.6711 & 0.9624 & 0.3175 & 0.049 & 0.834 & \textbf{3.1} & 500$^*$ \\
\addlinespace[3pt]
\bottomrule
\end{tabular}%
}
\flushleft
\footnotesize{$^*$ Note: The 500 realizations used for Mmin\_p and Mmin\_m are drawn from the remaining subset of the 15,000 fiducial simulations and are completely independent of the 14,500 realizations used for covariance estimation.}
\end{table*}
}

\newcommand{\tabnoisecorrected}{
\begin{table*}[t]
\centering
\caption{
The marginalized $1\sigma$ Fisher constraints derived from $P_{\rm h,0}$, $P_{\rm ZI-1,0}$, $P_{\rm ZI-10,0}$, $P_{\rm log,0}$, and the combined statistic
$P_{\rm ZI,0}=P_{\rm ZI-1,0}\oplus P_{\rm ZI-10,0}\oplus P_{\rm log,0}$. 
And $P_{\rm h,0}$ represents the halo power spectrum after subtracting the Poisson shot noise contribution.
For the ZI power spectra, values outside parentheses are obtained without noise correction, while values inside parentheses correspond to the noise-corrected results.
}
\label{tab:noise_corrected}
\small
\setlength{\tabcolsep}{9pt}
\renewcommand{\arraystretch}{1.15}
\begin{tabular}{lcccccc}
\toprule
Parameters 
& $P_{\rm h,0}$
& $P_{\rm ZI-1,0}$
& $P_{\rm ZI-10,0}$
& $P_{\rm log,0}$
& $P_{\rm ZI,0}$
& $P_{\rm h,0}\oplus P_{\rm ZI,0}$ \\
\midrule

$f_{\rm NL}^{\rm local}$ 
&146&143 (135)&147 (134)&148 (133)&116 (114)&101 (99)\\

$f_{\rm NL}^{\rm equil}$ 
&1452&1263 (631)&1339 (612)&1365 (601)&295 (257)&269 (232)\\

$f_{\rm NL}^{\rm ortho}$ 
&996&746 (465)&870 (492)&913 (497)&175 (180)&166 (160)\\

$M_\nu$ 
&0.313&0.307 (0.105)&0.333 (0.105)&0.341 (0.104)&0.113 (0.060)&0.092 (0.050)\\

$h$ 
&0.184&0.140 (0.072)&0.148 (0.075)&0.150 (0.075)&0.068 (0.043)&0.061 (0.036)\\

$n_{\rm s}$ 
&0.179&0.102 (0.059)&0.096 (0.057)&0.091 (0.056)&0.038 (0.031)&0.033 (0.027)\\

$\Omega_{\rm m}$ 
&0.033&0.028 (0.025)&0.028 (0.024)&0.028 (0.024)&0.016 (0.014)&0.013 (0.012)\\

$\Omega_{\rm b}$ 
&0.0151&0.0128 (0.0079)&0.0133 (0.0079)&0.0135 (0.0080)&0.0074 (0.0048)&0.0066 (0.0042)\\

$\sigma_8$ 
&0.024&0.091 (0.066)&0.090 (0.067)&0.088 (0.066)&0.026 (0.034)&0.015 (0.014)\\

$M_{\rm min}$
&0.512&0.551 (0.429)&0.609 (0.429)&0.630 (0.427)&0.219 (0.226)&0.140 (0.122)\\

\bottomrule
\end{tabular}
\end{table*}
}

\newcommand{\tabfishercompareall}{
\begin{table*}[t]
\centering
\caption{Comparison of Fisher constraints from the halo power spectrum, ZI power spectrum, marked power spectrum, and their combinations.}
\label{tab:fisher_compare_all}
\renewcommand{\arraystretch}{1.1}
\resizebox{\textwidth}{!}{
\begin{tabular}{lccccccccc}
\toprule
\addlinespace[1pt]
Parameters & $P_{\rm h,0}$ & $P_{\rm ZI,0}$ & $\sum M_{i,0}$ &
$P_{\rm h,0}\oplus P_{\rm ZI,0}$ &
$P_{\rm h,0}\oplus \sum M_{i,0}$ &
$P_{\rm h,0}\oplus P_{\rm ZI,0}\oplus \sum M_{i,0}$ &
$\frac{P_{\rm h,0}}{P_{\rm h,0}\oplus P_{\rm ZI,0}}$ &
$\frac{P_{\rm h,0}}{P_{\rm h,0}\oplus \sum M_{i,0}}$ &
$\frac{P_{\rm h,0}}{P_{\rm h,0}\oplus P_{\rm ZI,0}\oplus \sum M_{i,0}}$\\
\addlinespace[1pt]
\midrule
\addlinespace[2pt]
$f_{\rm NL}^{\rm local}$ &146&116&127&101&115&91&\textbf{1.4}&1.3&1.6\\
$f_{\rm NL}^{\rm equil}$&1452&295&276&269&230&191&5.4&\textbf{6.3}&7.6\\
$f_{\rm NL}^{\rm ortho}$&996&175&153&166&136&110&6.0&\textbf{7.3}&9.1\\
$M_\nu$&0.313&0.113&0.116&0.092&0.104&0.069&\textbf{3.4}&3.0&4.5\\
$h$&0.184&0.068&0.067&0.061&0.064&0.047&\textbf{3.0}&2.9&3.9\\
$n_{\rm s}$&0.179&0.038&0.045&0.033&0.042&0.027&\textbf{5.4}&4.2&6.7\\
$\Omega_{\rm m}$&0.033&0.016&0.017&0.013&0.016&0.010&\textbf{2.5}&2.0&3.3\\
$\Omega_{\rm b}$&0.0151&0.0074&0.0069&0.0066&0.0065&0.0051&\textbf{2.3}&\textbf{2.3}&2.9\\
$\sigma_8$&0.024&0.026&0.019&0.015&0.016&0.012&\textbf{1.7}&1.6&2.1\\
$M_{\rm min}$&0.512&0.219&0.190&0.140&0.167&0.111&\textbf{3.7}&3.1&4.6\\
\bottomrule
\end{tabular}}
\end{table*}
}

\usepackage{makecell}

\newcommand{\tabconvergence}{
\begin{table*}[t]
\centering
\caption{
The marginalized $1\sigma$ Fisher constraints derived from $P_{\rm h,0}$, $P_{\rm ZI,0}$, $\sum M_{i,0}$, and their combined statistics.
All values are extrapolated to $N_{\rm deriv}=5000$ by fitting the convergence curves measured from the 500 available simulations; no additional simulations are used for the extrapolation.
}
\label{tab:convergence}
\small 
\setlength{\tabcolsep}{9.5pt}
\renewcommand{\arraystretch}{1.15} 
\begin{tabular}{lcccccccc} 
\toprule
Parameters
& $P_{\rm h,0}$ & $P_{\rm ZI,0}$ & \makecell{$P_{\rm ZI,0}$\\[-2pt]{\scriptsize (noise-corrected)}} & $\sum M_{i,0}$ & $P_{\rm h,0}\oplus P_{\rm ZI,0}$ & \makecell{$P_{\rm h,0}\oplus P_{\rm ZI,0}$\\[-2pt]{\scriptsize (noise-corrected)}} & $P_{\rm h,0}\oplus\sum M_{i,0}$ & All \\
\midrule

$f_{\rm NL}^{\rm local}$
& 147
& 117
& 116
& 128
& 102
& 100
& 116
& 92 \\

$f_{\rm NL}^{\rm equil}$
& 1996
& 299
& 294
& 282
& 280
& 275
& 233
& 199 \\

$f_{\rm NL}^{\rm ortho}$
& 1243
& 177
& 183
& 154
& 169
& 162
& 138
& 111 \\

$M_\nu$
& 0.603
& 0.141
& 0.070
& 0.154
& 0.127
& 0.056
& 0.135
& 0.094 \\

$h$
& 0.425
& 0.069
& 0.044
& 0.078
& 0.062
& 0.038
& 0.074
& 0.051 \\

$n_{\rm s}$
& 0.209
& 0.039
& 0.032
& 0.045
& 0.034
& 0.027
& 0.043
& 0.027 \\

$\Omega_{\rm m}$
& 0.036
& 0.016
& 0.014
& 0.017
& 0.013
& 0.012
& 0.016
& 0.010 \\

$\Omega_{\rm b}$
& 0.0302
& 0.0077
& 0.0052
& 0.0075
& 0.0069
& 0.0046
& 0.0072
& 0.0055 \\

$\sigma_8$
& 0.027
& 0.027
& 0.037
& 0.020
& 0.015
& 0.014
& 0.016
& 0.012 \\

$M_{\rm min}$
& 0.564
& 0.221
& 0.229
& 0.192
& 0.143
& 0.124
& 0.169
& 0.113 \\

\bottomrule
\end{tabular}
\end{table*}
}